\documentclass[global]{svjour}
\usepackage{amsmath,amsfonts,amssymb}
\usepackage[sort&compress,numbers,square]{natbib}

\usepackage{xy}
    \xyoption{matrix}
    \xyoption{arrow}

\journalname{Communications in Mathematical Physics}

\title{Quantum Field Theory on Curved Backgrounds. I}

\subtitle{The Euclidean Functional Integral}

\author{Arthur Jaffe \and Gordon Ritter% etc
}                     % Do not remove
\institute{Harvard University\\
17 Oxford St.\\
Cambridge, MA 02138\\
\email{arthur\_jaffe@harvard.edu, ritter@post.harvard.edu}}
\date{August 16, 2006}
    \numberwithin{equation}{section}
    \numberwithin{theorem}{section}
    \numberwithin{conjecture}{section}
    \numberwithin{definition}{section}
    \numberwithin{example}{section}
    \numberwithin{remark}{section}
    \numberwithin{corollary}{section}
    \numberwithin{lemma}{section}

\newcommand{\lrp}[1]{\left( #1 \right)}
\newcommand{\lrpBig}[1]{\Big( #1 \Big)}

\newcommand{\lra}[1]{\left\langle #1 \right\rangle}
\newcommand{\abs}[1]{\left| #1 \right|}
\newcommand{\norm}[1]{\left\| #1 \right\|}

\newcommand{\cB}{\mathcal{ B}}
\newcommand{\cC}{\mathcal{ C}}
\newcommand{\cD}{\mathcal{ D}}
\newcommand{\cE}{\mathcal{ E}}

\newcommand{\cH}{\mathcal{ H}}
\def\H{\mathcal{H}}

\newcommand{\cK}{\mathcal{ K}}

\newcommand{\cN}{\mathcal{ N}}
\newcommand{\cO}{\mathcal{ O}}

\newcommand{\cP}{\mathcal{ P}}

\newcommand{\cT}{\mathcal{ T}}
\newcommand{\cU}{\mathcal{ U}}

\def\R{\mathbb{R}}

\def\C{\mathbb{C}}

\def\bbH{\mathbb{H}}

\newcommand\g{\mathfrak{g}}

\newcommand\fK{\mathfrak{K}}
\newcommand\fC{\mathfrak{C}}

\def\eps{\epsilon}
\def\vp{\varphi}

\def\th{\theta}

\def\la{\lambda}

\def\Om{\Omega}
\def\m{\mu}
\def\n{\nu}

\def\Th{\Theta}
\def\Ga{\Gamma}
\def\de{\delta}
\def\De{\Delta}

\def\d{\partial}

\newcommand\supp{\operatorname{supp}}

\newcommand\Iso{\operatorname{Iso}}

\newcommand\Diff{\operatorname{Diff}}

\newcommand{\Lie}{\operatorname{Lie}}

\def\<{\langle}
\def\>{\rangle}
\newcommand\del{\nabla}

\newcommand\To{\longrightarrow}

\newcommand{\dvol}{d\text{vol}}
\newcommand{\no}[1]{{\,: #1 :\,}}

\newcommand\Mapsto{\longmapsto}

\newcommand\te{\text}
\renewcommand\implies{\,\Rightarrow\,}
\newcommand\f{\frac}

\def\norm#1{\| #1 \|}
\def\abs#1{\left| #1 \right|}

\newcommand\scrg{\mathcal{G}}

\renewcommand{\iff}{\Leftrightarrow}

\newcommand{\ol}{\overline}

\newcommand\bea{\begin{eqnarray}}
\newcommand\eea{\end{eqnarray}}
\newcommand\beas{\begin{eqnarray*}}
\newcommand\eeas{\end{eqnarray*}}

\newcommand{\er}[1]{\eqref{#1}}
\newcommand{\bel}[1]{\begin{equation}\label{#1}}
\newcommand\ee{\end{equation}}
\newcommand\lab{\label}

\newcommand{\cale}{\mathcal{E}}
\newcommand{\calh}{\mathcal{H}}

\newcommand\HatOverSpace{{\hat{\phantom{a}}}}
\newcommand{\Quantization}[1]{#1\HatOverSpace}

\newcommand{\Gspace}{G_{\te{space}}}
\newcommand{\gsp}{\g_{\te{sp}}}
\newcommand{\hyp}[1]{{\mathbb{H}^{\,{#1}}}}

\newcommand\vx{{\vec x}}
\newcommand\vy{{\vec y}}
\newcommand\vpsqr{{\vec p}{\,}^2}

\newcommand{\exph}[1]{\left\langle #1 \right\rangle_{\mathcal{H}}}
\newcommand{\expe}[1]{\left\langle #1 \right\rangle_{\mathcal{E}}}

\newfont{\boldit}{cmbxti10}

\begin{document}

\maketitle

\hyphenation{Oster-walder}

\begin{abstract}
We give a mathematical construction of Euclidean quantum field theory
on certain curved backgrounds. We focus on generalizing
Osterwalder Schrader quantization, as these methods have proved
useful to establish estimates for interacting fields on flat
space-times. In this picture, a static Killing vector generates translations in Euclidean time, and physical positivity is
played by positivity under reflection of Euclidean time. We discuss
the quantization of flows which correspond to classical space-time symmetries,
and give a general set of conditions which imply that broad classes
of operators in the classical picture give rise to well-defined
operators on the quantum-field Hilbert space. In particular,
Killing fields on spatial sections give rise to unitary groups on
the quantum-field Hilbert space, and corresponding densely-defined self-adjoint
generators. We construct the Schr\"odinger
representation using a method which involves localizing certain
integrals over the full manifold to integrals over a codimension-one
submanifold. This method is called sharp-time localization, and implies
reflection positivity.
\end{abstract}

\maketitle

%%%%%%%%%%%%%%%%%%%%%%%%%%%%%%%%%%%%%%%%%%%%%%%%%%%%%%%%%%%%
\section*{Introduction}
%%%%%%%%%%%%%%%%%%%%%%%%%%%%%%%%%%%%%%%%%%%%%%%%%%%%%%%%%%%%

The present article presents a construction of a Euclidean quantum
field theory on time-independent, curved backgrounds. Earlier work
on field theories on curved space-time  (Kay \cite{Kay:1978}, Dimock
\cite{Dimock84}, Bros et al. \cite{BEM02}) uses
real-time/Lorentzian signature and algebraic
techniques reminiscent of $\cP(\varphi)_2$ theory from the
Hamiltonian point of view \cite{Expositions}.  In contrast, the
present treatment uses the Euclidean functional integral \cite{GJ}
and Osterwalder-Schrader quantization \cite{OS,OS2}.  Experience
with constructive field theory on $\R^d$ shows that the Euclidean
functional integral provides a powerful tool, so it is interesting
also to develop Euclidean functional integral methods for manifolds.

Euclidean methods are known to be useful in the study of black
holes, and a standard strategy for studying black hole (BH)
thermodynamics is to analytically continue time in the BH metric
\cite{Carlip:95a}. The present paper implies a mathematical
construction of scalar fields on any static, Euclidean black hole
background. The applicability of the Osterwalder-Schrader
quantization procedure to curved space depends on unitarity of the
time translation group and the time reflection map which we prove
(theorem \ref{thm:kernel}). The Osterwalder-Schrader construction
has universal applicability; it contains the Euclidean functional
integral associated with scalar boson fields, a generalization of
the Berezin integral for fermions, and a further generalization
for gauge fields \cite{Ashtekar}. It also appears valid for fields
on Riemann surfaces \cite{Jaffe:1989zv}, conformal field theory
\cite{Gawedzki}, and may be applicable to string theory. The present paper
extends this construction to models on curved backgrounds.

Our paper has many relations with other work. Wald
\cite{Wald79} studied metrics with Euclidean signature, although he
treated the functional integral from a physical rather than a
mathematical point of view. Brunetti et al \cite{Verch:2003a} developed the
algebraic approach (Haag-Kastler theory) for curved space-times and
generalized the work of Dimock \cite{DimockAlgebras}.  They describe
covariant functors between the category of globally hyperbolic
spacetimes with isometric embeddings, and the category of
$\ast$-algebras with unital injective $\ast$-monomorphisms.

The examples studied in this paper---scalar quantum field theories
on static space-times---have physical relevance. A first
approximation to a full quantum theory (involving the gravitational
field as well as scalar fields) arises from treating the sources of
the gravitational field classically and independently of the
dynamics of the quantized scalar fields \cite{BirrellDavies}. The
weakness of gravitational interactions, compared with elementary
particle interactions of the standard model, leads one to believe
that this approximation is reasonable. It exhibits nontrivial
physical effects which are not present for the scalar field on a
flat spacetime, such as the Hawking effect \citep{Hawking75} or the
Fulling-Unruh effect \citep{Unruh:1976db}. Density perturbations in
the cosmic microwave background (CMB) are calculated using scalar
field theory on certain curved backgrounds \cite{LythRiotto}.
Further, Witten \cite{Witten} used quantum field theory on Euclidean anti-de
Sitter space in the context of the AdS/CFT correspondence
\cite{Malda,GKP}.

Some of the methods discussed here in Section \ref{chap:quantization} have
been developed for the flat case in lecture courses; see \cite{Jaffe:2005a}.

\subsection*{Notation and conventions}

We use notation, wherever possible, compatible with standard
references on relativity \citep{WaldGR} and quantum field theory
\citep{GJ}. We use Latin indices $a,b = 0 \ldots d-1$ for
spacetime indices, reserving Greek indices $\mu, \nu = 1 \ldots
d-1$ for spatial directions. We include in our definition of
`Riemannian manifold' that the underlying topological space must
be paracompact (every open cover has a locally finite open
refinement) and connected. The notation $L^2(M)$ is used when $M$ is a
$C^\infty$ Riemannian manifold, and implicitly refers to the Riemannian volume
measure on $M$, which we sometimes denote by $\dvol$. Also $\cU(\cH)$ denotes
the group of unitary operators on $\cH$. Let $G = I(M) = \Iso(M)$ denote the
isometry group, while $\fK$ is its Lie algebra, the global Killing fields. For
$\psi$ a smooth map between manifolds, we use $\psi^*$ to denote the pullback
operator $(\psi^*f)(p) = f(\psi(p))$. The notation $\De = \De_M$ means the
Laplace operator for the Riemannian metric on $M$.

\section{Reflection Positivity}
%%%%%%%%%%%%%%%%%%%%%%%%%%%%%%%%%%%%%%%%%%%%%%%%%%%%%%%%%%%%
\subsection{Analytic continuation} \label{sec:analytic}
%%%%%%%%%%%%%%%%%%%%%%%%%%%%%%%%%%%%%%%%%%%%%%%%%%%%%%%%%%%%

The Euclidean approach to quantum field theory on a curved
background has advantages since elliptic operators are easier to
deal with than hyperbolic operators. To obtain physically meaningful
results one must presumably perform the analytic continuation back
to real time. In general, Lorentzian spacetimes of interest may not
be sections of 4-dimensional complex manifolds which also have
Riemannian sections, and even if they are, the Riemannian section
need not be unique. Thus, the general picture of extracting physics
from the Euclidean approach is a difficult one where further
investigation is needed.

Fortunately, for the class of spacetimes treated in the present
paper (static spacetimes), the embedding within a complex 4-manifold
with a Euclidean section is guaranteed, and in such a way that
Einstein's equation is preserved \cite{Chrusciel:2004yv}.

\subsection{Time reflection} \label{sec:time-refl}

Reflection in Euclidean time plays
a fundamental role in Euclidean quantum field theory, as shown by Osterwalder
and Schrader \cite{OS,OS2}.

\begin{definition}[Time reflection] \label{time-reflection}
Let $M$ be a Riemannian manifold.
A {\bf time reflection} $\th : M \to M$ is an isometric
involution which fixes pointwise a smooth codimension-one
hypersurface $\Sigma$. This means that $\th \in \Iso(M)$,
$\th^2 = 1$ and $\th(x) = x$ for all $x \in \Sigma$.
\end{definition}

We now discuss time reflection for static manifolds, which is the example that
we will study in this paper.

\begin{example}[Static manifolds] \label{ex-theta-a}
Suppose there exists a globally defined, static Killing field $\xi$.
Fix a hypersurface $\Sigma \subset M$ to which $\xi$ is orthogonal.
Define a global function $t : M \to \R$ by setting $t = 0$ on $\Sigma$, and otherwise define $t(p)$ to be the unique
number $t$ such that $\phi_t(x) = p$ for some $x \in \Sigma$, where
$\{ \phi_t \}$ is the one-parameter group of isometries determined by
$\xi$. Finally, define $\th$ to map a point $p \in M$ to the
corresponding point on the same $\xi$-trajectory but with
$t(\th(p)) = -t(p)$. This defines a decomposition
\bel{M}
    M = \Om_- \cup \Sigma \cup \Om_+, \quad \th\Om_\pm = \Om_\mp ,
    \quad \th \Sigma = \Sigma .
\ee
\end{example}

In past work \cite{Jaffe:1989zv}, we have considered time-reflection maps
which fall outside the bounds of example \ref{ex-theta-a} (\cite{Jaffe:1989zv}
applies to compact Riemann surfaces, which cannot support Killing fields), but
we will not do so here.

The time-reflection map given by a hypersurface-orthogonal Killing
field is not unique, but depends on a choice of the
\emph{initial hypersurface}, which we fix.
The initial hypersurface will be used to define time-zero fields.
Reflection of the Euclidean time coordinate $t \to -t$
analytically continues to Hermitian conjugation of $e^{-itH}$.

%%%%%%%%%%%%%%%%%%%%%%%%%%%%%%%%%%%%%%%%%%%%%%%%%%%%%%%%%%%%
\subsection{Fundamental assumptions}\label{sec:fundamentals}
%%%%%%%%%%%%%%%%%%%%%%%%%%%%%%%%%%%%%%%%%%%%%%%%%%%%%%%%%%%%

Let $C = (- \De + m^2)^{-1}$ be the resolvent of the Laplacian, also called the
\emph{free covariance}, where $m^2 > 0$. Then $C$ is a bounded self-adjoint
operator on $L^2(M)$. For each $s \in \R$, the Sobolev space $H_s(M)$ is a real
Hilbert space, which can be defined as completion of $\cC^{\infty}(M)$ in the
norm
\bel{SobolevNorm}
	\|f \|^2_s = \< f, C^{-s} f \> .
\ee

We work with test functions in $H_{-1}(M)$. This is a convenient
choice for several reasons: the norm \er{SobolevNorm} with $s = -1$ is related
in a simple way to the free covariance, and further, Dimock
\cite{Dimock:2003an} has given an elegant proof of reflection
positivity for Sobolev test functions. Another motivation is as
follows. Suppose we wish to prove that $\vp(h)$ is a bounded
perturbation of the free Hamiltonian $H_0$ for a scalar field on
$\R^d$. The first-order perturbation is
\bel{eqn:first-order}
    -\<\Omega_1, H_0 \Omega_1\> =
    -\f12 \int \f{|\hat h(\vec p)|^2}{\omega(\vec p)^2} d\vec p
\ee
where we used $\Omega_1 = - H_0^{-1} \vp(h) \Omega$. Existence
of \er{eqn:first-order} is equivalent to $h \in H_{-1}(\R^d)$, so
this is a natural condition for test functions. Therefore we choose $H_{-1}(M)$
for the generalization to curved manifolds.

The Sobolev spaces give rise to a natural rigging, or \emph{Gelfand triple},
and various associated Gaussian measures \cite{GV,Simon}. The inclusion $H_s
\hookrightarrow H_{s+k}$ for $k > 0$ is Hilbert-Schmidt, so the spaces
\[
	H_\infty \equiv \bigcap_{s =1}^\infty H_s(M)
	\ \ \subset \ \
	H_{-1}(M)
	\ \ \subset \ \
	\bigcup_{s=-1}^{-\infty} H_s(M) \equiv H_{-\infty}
\]
form a Gelfand triple, and $H_\infty$ is a nuclear space. There is a unique
Gaussian measure $\mu$ defined on the dual $H_{-\infty}$ with  covariance $C$. This means that
\[
	S(f) \equiv
	\int_{H_{-\infty}}
	e^{i \Phi(f)} \, d\mu (\Phi )
	=
	e^{-\f12 \< f, Cf\>}, \quad f\in H_\infty .
\]
Define
\[
	\cE\  := \  L^2(H_{-\infty}, \mu ).
\]
The space $\cE$ is unitarily equivalent to
Euclidean Fock space over $H_{-1}(M)$ (see for example \cite[Theorem I.11]{Simon}). The algebra generated by monomials of
the form $\Phi(f_1) \ldots \Phi(f_n)$ is dense in $\cE$.
This is a special case of a general construction discussed
in the reference.

\begin{definition}[Standard domain] \label{def:standard-domain}
For an open set $\Omega \subseteq M$,
the \emph{standard domain} in $\cE$ corresponding to $\Om$ is:
\[
    E_\Om = \mathrm{span}\{ e^{i \Phi(f)} : f \in H_{-1}(M),\, \supp(f)
\subset \Om\}.
\]
Let $\cE_\Om$ denote the closure in $\cE$ of $E_\Om$.
\end{definition}

Definition \ref{def:standard-domain} refers to subspaces of $\cE$
generated by functions supported in an open set. This includes empty
products, so $1 \in \cE_\Om$ for any $\Om$. Of particular importance
for Euclidean field theory is the positive-time subspace
\[
    \cE_+ \ :=\  \cE_{\Om_+},
\]
where the notation $\Om_+$ refers to the decomposition
\er{M}. A linear operator on $\cE$ which maps $\cE_+\to \cE_+$ is said to be
\emph{positive-time invariant}.

%%%%%%%%%%%%%%%%%%%%%%%%%%%%%%%%%%%%%%%%%%%%%%%%%%%%%%%%%%%%
\subsection{Operator induced by a diffeomorphism}
\label{sec:operator-induced}
%%%%%%%%%%%%%%%%%%%%%%%%%%%%%%%%%%%%%%%%%%%%%%%%%%%%%%%%%%%%

We will consider the effect which diffeomorphisms of the underlying
spacetime manifold have on the Hilbert space operators which arise
in the quantization of a classical field theory. For $f \in
C^\infty(M)$ and $\psi : M \to M$ a diffeomorphism, define
\bel{Superscript}
    f^\psi \ \equiv\  \psi_* f = (\psi^{-1})^* f = f \circ \psi^{-1}.
\ee
The reason for using $\psi^{-1}$ here is so that Definition \ref{def:Gamma} gives a group representation. 

\begin{definition}[Induced operator] \label{def:Gamma}
Let $\psi$ be a diffeomorphism, and
$A(\Phi) = \Phi(f_1) \ldots \Phi(f_n) \in \cE$ a monomial. Define
\bel{WickOrderedGamma}
    \Gamma(\psi) \no{A}
    \ \equiv\
    \no{ \Phi({f_1}^\psi) \cdots \Phi({f_n}^\psi) }\, .
\ee
This extends linearly to a dense domain in
$\cE$. We refer to $\Gamma(\psi)$ as the operator
{\bf induced by} the diffeomorphism $\psi$.
\end{definition}

Note that if $\psi$ is an isometry, then \er{WickOrderedGamma}
is equivalent to the definition $\Gamma(\psi) A \ \equiv\ \Phi({f_1}^\psi) \ldots \Phi({f_n}^\psi)$ without Wick ordering, as follows from
\er{equivalence1} below.

The induced operators $\Ga(\psi)$ are \emph{not} necessarily bounded on $\cE$.
In fact, for a general diffeomorphism $\psi$, the operator $\psi^*$ may fail to
be bounded on $L^2(M)$ or $H_{-1}(M)$. If the Jacobian $|d\psi|$ satisfies
uniform upper and
lower bounds, i.e.
\bel{diffeo-bound}
    (\exists \, c_1, c_2 > 0) \ \
    c_1 < \sup_{x \in M} |d\psi_x| <  c_2.
\ee
then $(\psi^{-1})^*$ is bounded on $L^2(M)$, but
$\Ga(\psi)$ may still be unbounded on $\cE$, because the operator norm of
$\Ga(\psi)$ on the degree-$n$ subspace of $\cE$ may fail to have a limit as $n
\to \infty$. In this situation, $\Ga(\psi)$ is to be regarded as a
densely-defined unbounded operator whose domain includes all finite particle
vectors.

If $(\psi^{-1})^*$ is a contraction on $H_{-1}(M)$, then $\Ga(\psi)$ is a
contraction on $\cE$ (in particular, bounded).
A special case of this is $\psi \in \Iso(M)$, which implies that
$\Ga(\psi)$ is unitary and $\norm{\Ga(\psi)}_{\cE} = 1$.

\begin{lemma}[Naturalness property]\label{lem:isometry1}
Let $\psi : M \to M$ be a diffeomorphism, and consider the pullback
$\psi^*$ acting on $L^2(M)$, with its Hermitian adjoint $(\psi^*)^\dag$.
Then
\bel{volp}
	\det(d\psi) = 1
	\ \iff\
	(\psi^*)^\dag  =  (\psi^{-1})^*
	\ \iff\
	\psi \te{ is volume-preserving }.
\ee
Furthermore,
\bel{equivalence1}
	\psi \in \Iso(M)
	\ \iff\
	\Gamma(\psi) \in \cU(\cE)
	\ \iff\
	[\psi^*, \De] = 0
	\ \iff\
	[\psi^*, C] = 0 .
\ee
\end{lemma}
The last part of \er{equivalence1} follows from
\cite[Theorem III.6.5]{Kato}, while the rest of the statements in \er{volp} and
\er{equivalence1} are standard calculations. It follows that $\Ga$ restricts to
a unitary representation of $G = \Iso(M)$ on $\cE$.

For an open set $\Om \subset M$, define
\[
    \Iso(M,\Om) = \{ \psi \in \Iso(M) : \psi(\Om) \subset \Om
    \},
\]
and similarly $\Diff(M,\Om)$. These are not subgroups of
$\Diff(M)$ but they are semigroups under composition. If $\psi \in
\Diff(M,\Om)$ we say $\psi$ preserves $\Om$.

\begin{lemma}[Presheaf property]  \label{lem:presheaf}
Let $\psi : U \to V$ be a diffeomorphism, where $U, V$ are open sets in $M$.
Let $E_U, E_V$ be the corresponding standard domains (cf. Definition
\ref{def:standard-domain}). Then
\[
    \Ga(\psi)E_U = E_V.
\]
In particular, if $\psi : M \to M$ preserves $\Om \subset M$, then $\Ga(\psi)$
preserves the corresponding subspace $\cE_{\Om} \subset \cE$.
\end{lemma}

For maps $\psi : U \to V$ which are subset inclusions $U \subseteq
V$, lemma \ref{lem:presheaf} asserts that the association $U \to
\cE_U$ is a \emph{presheaf}. It also follows from lemma
\ref{lem:presheaf} that the mappings $U \to \cE_U$ and $\psi \to
\Ga(\psi)$ define a covariant functor from the category of open
subsets of $M$ with invertible, smooth maps between them into the
category of Hilbert spaces and densely defined operators.

Lemma \ref{lem:presheaf} implies that if $\psi(\Om_+) \subset \Om_+$ then
$\Ga(\psi)$ is positive-time invariant. This is  necessary but not sufficient
for $\Ga(\psi)$ to have a quantization. A sufficient condition is that $\Ga(\psi)$ and $\Th \Ga(\psi)^\dag \Th$ both preserve $\cE_+$, where $\Th = \Ga(\th)$, as shown by
Theorem \ref{Prop:NullInvariance}.

%%%%%%%%%%%%%%%%%%%%%%%%%%%%%%%%%%%%%%%%%%%%%%%%%%%%%%%%%%%%
\subsection{Continuity results}
\label{sec:strongcont}
%%%%%%%%%%%%%%%%%%%%%%%%%%%%%%%%%%%%%%%%%%%%%%%%%%%%%%%%%%%%

\begin{lemma}[Sobolev continuity]\label{topology-on-cE} For the free covariance
$C = (-\De + m^2)^{-1}$,
\[
    \{f_1, \ldots, f_n\} \ \longmapsto \
    A(\Phi) := \Phi(f_1) \ldots \Phi(f_n) \in \cE
\]
is continuous $(H_{-1})^n \to \cE$, where we take
the product of the Sobolev topologies on $(H_{-1})^n$.
\end{lemma}

\begin{proof}
Since $\Phi$ is linear, it is sufficient to show that
$\norm{A(\Phi)}_\cale$ is bounded by
$\te{const.}\prod_i\norm{f_i}_{-1}$. As a consequence of the
Gaussian property of the measure $d\mu_C$, one needs only bound the
linear case.  But
    \bel{eq:sobolev-norm}
        \norm{\Phi(f)}_{\cale}
        = \Big| \int \lrp{\Phi(f) \Phi(f)} \, d\mu_C \Big|^{1/2}
        = \norm{f}_{-1} \,.
    \ee
\end{proof}

\begin{theorem}[Strong continuity] \label{thm:strongcont}
Let $\{\psi_n\}$ be a sequence of orientation-preserving isometries
which converge to $\psi$ in the compact-open topology. Then
$\Ga(\psi_n) \to \Ga(\psi)$ in the strong operator topology on $\cB(\cE)$.
\end{theorem}

The proof of theorem \ref{thm:strongcont} follows standard
arguments in analysis. Let us give a sense of how it is to be used. If all the
elements of a certain one-parameter group of isometries $\psi_t$ are such that
$\Ga(\psi_t)$ have bounded quantizations, then $t \to \hat \Ga(\psi_t)$ defines
a one-parameter group of operators on $\cH$ (the quantum-field Hilbert space).
In this situation, Theorem \ref{thm:strongcont} justifies the
application of Stone's theorem. This picture is to be developed in
section \ref{chap:quantization}.

%%%%%%%%%%%%%%%%%%%%%%%%%%%%%%%%%%%%%%%%%%%%%%%%%%%%%%%%%%%%
\subsection{Reflection positivity}\label{sec:RP}
%%%%%%%%%%%%%%%%%%%%%%%%%%%%%%%%%%%%%%%%%%%%%%%%%%%%%%%%%%%%

\begin{definition}\label{def:RP}
With $\th$ as in Definition \ref{time-reflection}, let $\Th =
\Gamma(\theta)$ be the induced reflection on $\cE$. A measure $\mu$ on $H_{-\infty}$ is
said to be \emph{reflection positive} if
\bel{action}
    \int \overline{ \Theta(F)}\, F \ d\mu  \geq 0
    \ \ \te{for all} \ \ F \in \cE_+ \, .
\ee
A bounded operator $T$ on $L^2(M)$ is said to be \emph{reflection positive}
if
\bel{OperatorRP}
   \supp f \subseteq \Om_+ \ \Rightarrow\ \< f, \th T f\>_{L^2(M)} \geq 0.
\ee
\end{definition}

Reflection positivity for the measure $\mu$
is equivalent to the following inequality for
operators on $\cE = L^2(d\mu)$:
\[
    0 \leq \Pi_+ \Th \Pi_+
\]
where $\Pi_+ : \cE \to \cE_+$ is the canonical projection.

A Gaussian measure with mean zero and covariance $C$ is reflection
positive iff $C$ is reflection positive in the operator
sense, eqn.~\er{OperatorRP}. An equivalent condition is that for any finite
sequence $\{f_i\}$ of real functions supported in $\Om_+$, the
matrix $M_{ij} = \exp {\<f_i,\, \th C f_{j}\>}$ has no negative eigenvalues.

For Riemannian manifolds which possess an isometric involution whose fixed-point set has codimension one, there is a simple
potential-theoretic proof of reflection positivity \cite{DDD86}. The relation between reflection positivity and operator
monotonicity under change of boundary conditions for the Laplacian was
discovered in \cite{NoteOnRP}.
A different proof of reflection positivity
on curved spaces was given by Dimock \cite{Dimock:2003an}, based on
Nelson's proof using the Markov property \cite{Nelson}. We give a third proof
later in this paper based on our sharp-time localization theorem. The result is summarized as follows.

\begin{theorem}[Reflection positivity] \label{thm:rp}
Let $M$ be a Riemannian manifold with a time reflection as in Definition \ref{time-reflection}. Then the covariance $C = (-\Delta + m^2)^{-1}$ and its associated Gaussian measure are reflection positive.
\end{theorem}

\section{Osterwalder-Schrader Quantization and the Feynman-Kac Formula}
\label{chap:quantization}

The Osterwalder-Schrader construction is a
standard feature of quantum field theory.  It begins with a
``classical'' Euclidean Hilbert space $\cE$ and leads to the
construction of a Hilbert space $\cH=\Pi\cE_+$, which is the
projection $\Pi$ of the Euclidean space $\cE_+$.  It also yields a
quantization map $T\mapsto \hat T$ from a classical operator $T$ on $\cE$ to a
quantized operator $\hat T$ acting on $\cH$.  In this section we review this
construction, dwelling on the
quantization of bounded operators $T$ on $\cE$ that may yield a
bounded or an unbounded quantization $\hat T$, as well as the
quantization of an unbounded operator $T$ on $\cE$.   We give a
variation of the previously unpublished treatment in
\cite{Jaffe:2005a}, adapted to curved space-time.

\subsection{The Hilbert space}

Define a bilinear form $(A,B)$ on $\cE_+$ by
\bel{bform}
    ( A, B ) = \< \Th A, B\>_{\cE} \quad
    \te{ for } \quad A, B \in \cE_+ \, .
\ee
Using self-adjointness of $\Th$ on $\cE$, one can show that
this form is sesquilinear,
\bel{sesqui}
    (B,A) = \int \overline{\Th B}\, A\ d\mu
    =
    \lrpBig{\int B\, \overline{\Th A}\, d\mu}^*
        =
    \overline{(A,B)} \, .
\ee
If $\th$ is not an isometry,
then $\Th$ is non-unitary in which case Osterwalder-Schrader
quantization is not possible. Therefore, it is essential that $\th
\in \Iso(M)$. The form \er{bform} is degenerate, and has an
infinite-dimensional kernel which we denote $\cN$. Therefore
\er{bform} determines a nondegenerate inner product $\< \, ,
\,\>_{\cH}$ on $\cE_+ / \cN$, making the latter a pre-Hilbert space.

\begin{definition}[Hilbert space]\label{def:hilbertspace}
The (Osterwalder-Schrader) physical Hilbert space $\H$
is the completion of $\cE_+ /\cN$,
with inner product $\< \, , \,\>_{\cH}$.  Let $\Pi : \cE_+ \to
\cH$ denote the natural quotient map, a contraction mapping from  $A \in \cE_+$ to $\hat A := \Pi A$. There is an exact sequence:
\[
    \xymatrix{
    0 \ar[r] & \cN\ \ar[r]^{\te{incl.}} & \cE_+ \ar@{->>}[r]^{\Pi} &
    \cH \ar[r] & 0
    } \, .
\]
\end{definition}

%%%%%%%%%%%%%%%%%%%%%%%%%%%%%%%%%%%%%%%%%%%%%%%%%%%%%%%%%%%%
\subsection{Quantization of operators}\label{sec:general-quant}
%%%%%%%%%%%%%%%%%%%%%%%%%%%%%%%%%%%%%%%%%%%%%%%%%%%%%%%%%%%%

Assume that $T$ is a densely defined, closable operator on $\cE$ with domain
$\cD\subset\cE$. Define $T^+ := \Theta T^* \Theta$, and assume there exists a
subdomain $\cD_0\subset\cD\cap\cE_+$ on which $T^+$ is defined and
for which both
\bel{QuantizationCondition}
        T{:}\ \cD_0 \rightarrow \cE_+,
        \  \text{and}\ \
        T^+{:}\ \cD_0 \rightarrow \cE_+\;.
    \ee

\begin{theorem}[Condition for quantization] \label{Prop:NullInvariance}
Assume that $\hat\cD_0 := \Pi(\cD_0)$ is dense in $\cH$. Condition
\eqref{QuantizationCondition} ensures that $T$ has a quantization $\hat T$ with
domain $\hat\cD_0$. Furthermore $\hat T^*$ is defined, $\hat T$ has a closure,
and on $\hat\cD_0$, we have:
\bel{AdjointOnH}
	{\hat T }^* = \widehat{T^+} \, .
\ee
\end{theorem}

\begin{proof}
First, we check that $\hat T$ is well-defined. Suppose
$A\in\cN \cap \cD_0$. Let $B \in \cE_+$ range over a set of vectors
in the domain of $\Th T^*\Th$ such that the image of this set under
$\Pi$ is dense in $\cH$. Then
 \[
        0 = \< \lrp{\Th T^* \Th B}\HatOverSpace, \hat A \>_{\cH}
        = \< T^* \Th B, A \>_{\cE}
        = \< \Th B, T A \>_{\cE}
        = \< \hat B ,\widehat{TA} \>_{\cH}\;.
    \]
Thus $T A\in\cN$, and hence $T$ is well-defined on $\cD_0 / \cD_0 \cap \cN$.
To check \er{AdjointOnH} is a routine calculation.
%To check \er{AdjointOnH}, we have $\< \hat T \hat \psi, \hat\phi\>_\cH
%= \< \Th T \psi, \phi\>_\cE = \< \psi, T^* \Th \phi\>_\cE = \< \Th \psi, T^+
%\phi\>_\cE = \< \hat\psi, \widehat{T^+} \phi\>_\cH$.
\end{proof}

The main content of Theorem \ref{Prop:NullInvariance} can be
expressed as a commutative diagram. For bounded transformations, Theorem \ref{Prop:NullInvariance} simply means that if $T : \cE_+ \to \cE_+$ and the
dotted arrow in the following diagram is well-defined, then so are the two
solid arrows:
\[
    \xymatrix{
    0 \ar[r] & \cN \ar[r]^{\te{incl.}} \ar[d]^{T} &
    \cE_+ \ar[r]^{\Pi} \ar@{..>}[d]^{\Th T^* \Th} &
    \cH \ar[d]^{\hat T} \ar[r] & 0
    \\
    0 \ar[r] & \cN \ar[r]^{\te{incl.}} & \cE_+ \ar[r]^{\Pi} & \cH \ar[r] & 0
    }
\]

\begin{lemma}[Contraction property] \label{prop:contraction}
Let $T$ be a bounded transformation on $\cale$ such that $T$ and $\Th T^* \Th$
each preserve $\cE_+$. Then $\hat T$ is a bounded transformation on $\cH$ and
        \bel{MFNormBound}
            \norm{\hat{T}}_\calh
            \le \norm{T}_\cale\;.
        \ee
\end{lemma}

\begin{proof}
This proceeds by the multiple reflection method \cite{GJ}.
\end{proof}

We now discuss some examples of operators satisfying the
hypotheses of Theorem \ref{Prop:NullInvariance}.

\begin{theorem}[Self-adjointness]
\label{thm:sa} Let $U$ be unitary on $\cE$, and $U(\cE_+) \subset
\cE_+$. If $U^{-1} \Th = \Th U$ then $U$ admits a quantization $\hat
U$ and $\hat U$ is self-adjoint. (Do not assume $U^{-1}$
preserves $\cE_+$).
\end{theorem}

\begin{proof}
The operator $\Th U^* \Th = \Th^2 U = U$ preserves $\cE_+$, so Theorem
\ref{Prop:NullInvariance} $\Rightarrow$ $U$ has a quantization $\hat U$.
Self-adjointness of $\hat U$ follows from eqn.~\er{AdjointOnH}.
\end{proof}

\begin{theorem}[Unitarity]
\label{thm:unit} Let $U$ be unitary on $\cE$, and $U^{\pm 1}(\cE_+) \subset
\cE_+$. If $[U, \Th] = 0$ then $U$ admits a quantization $\hat U$ and $\hat U$
is unitary.
\end{theorem}

\begin{proof}
The operator $\Th U^* \Th = U^* = U^{-1}$ preserves $\cE_+$ by assumption, so $U$ has a quantization.
Also, $\Th (U^{-1})^* \Th = U$ preserves $\cE_+$, so $U^{-1}$ also
has a quantization. Obviously, the quantization of $U^{-1}$ is the
inverse of $\hat U$. Eqn.~\er{AdjointOnH} implies that the
adjoint of $\hat U$ is the quantization of $\Th U^* \Th = U^* =
U^{-1}$.
\end{proof}

Examples of operators satisfying the conditions of Theorems
\ref{thm:sa} and \ref{thm:unit} come naturally from isometries on
$M$ with special properties. We now discuss two classes of
isometries, which give rise to self-adjoint and unitary operators as above.

\begin{example}[Reflected Isometries] \label{ex:reflectediso}
An element $\psi \in \Iso(M)$ is said to be a {\bf reflected}
isometry if
    \bel{eqn:refliso}
      \psi^{-1} \circ \th = \th \circ \psi \, .
    \ee
If additionally $\psi(\Om_+) \subseteq \Om_+$ then
Theorem \ref{thm:sa} implies that $\hat \Ga(\psi) : \cH \to
\cH$ exists and is self-adjoint. If $\psi$ satisfies
\er{eqn:refliso} then so does $\psi^{-1}$; hence
if $\psi^{-1}(\Om_+) \subseteq \Om_+$, then
$\Ga(\psi^{-1})$ has a quantization and $\hat \Ga(\psi^{-1})$ is the
inverse of $\hat \Ga(\psi)$.
\end{example}

\begin{example}[Reflection-Invariant Isometries] \label{ex:reflection-invariant}
A {\bf reflection-invariant} isometry is an element $\psi \in
\Iso(M)$ that commutes with time-reflection, $\psi\th = \th\psi$.
It follows that $[\Ga(\psi), \Th] = 0$.
If $\psi$ and $\psi^{-1}$ both preserve $\Om_+$ then $\Ga(\psi^{\pm 1}) \cE_+ \subset \cE_+$, and Theorem \ref{thm:unit} implies that $\hat\Ga(\psi) : \cH \to \cH$ is unitary. The set of reflection-invariant isometries form a subgroup of the full isometry group.
\end{example}

%%%%%%%%%%%%%%%%%%%%%%%%%%%%%%%%%%%%%%%%%%%%%%%%%%%%%%%%%%%%
\subsection{Quantization domains}\label{sec:quantizationdomains}
%%%%%%%%%%%%%%%%%%%%%%%%%%%%%%%%%%%%%%%%%%%%%%%%%%%%%%%%%%%%

Quantization domains are subsets of
$\Om_+$ which give rise to dense domains in $\cH$ after quantization.
This is important for the analysis of unbounded operators on $\cH$. For example, an isometry which satisfies \er{eqn:refliso} may only map a proper subset $\cO \subset \Om_+$ into $\Om_+$, and in this case $\Ga(\psi)$ is only defined on a non-dense subdomain of $\cE_+$. If $\cO$ is a quantization domain, then $\Pi\cE_\cO$ may still be dense in $\cH$, and can serve as a domain of definition for $\hat\Gamma(\psi)$.

\begin{definition} \label{def:quantizationdom}
A {\bf quantization domain} is a subspace $\Om \subset \Om_+$ with
the property that $\Pi\lrp{\cE_\Om}$ is dense in $\cH$.
\end{definition}

\begin{example}
Perhaps the simplest quantization domain is a half-space lying at
times greater than $T>0$,
\bel{HalfSpace}
        \cO_{+,T}= \left\{ x\in\R^d : x_0 > T \right\} .
\ee
Let $\cD_{+,T} = E_{\cO_{+,T}} = \Gamma(\psi_{T})E_+$ where
$\psi_T(x,t) = (x,t+T)$; then $\Pi(\cD_{+,T})$ is dense in $\cH$, as follows from Theorem \ref{thm:generalQDs}. 
\end{example}

Theorem \ref{thm:generalQDs} generalizes \er{HalfSpace} to
curved spacetimes, and also allows one to replace the simple
half-space $\cO_{+,T}$ with a more general connected subset of
$\Om_+$.

\begin{theorem}[Construction of quantization domains] \label{thm:generalQDs}
Suppose $\psi \in \Iso(M,\Om_+)$, i.e. $\cO := \psi(\Om_+)
\subset \Om_+$. If $[\Ga(\psi), \Th] = 0$ or $\Ga(\psi) \Th = \Th
\Ga(\psi^{-1})$ (i.e. $\psi$ is reflection-invariant or reflected) then $\cO$ is a quantization domain.
\end{theorem}

\begin{proof}
By lemma \ref{lem:presheaf}, we have
\bel{EU}
    E_\cO = \Ga(\psi) E_+ \, .
\ee
Let $\hat C \in \cH$ be orthogonal to every vector $\hat A \in
\Pi(\cE_\cO)$. Choose $B \in \cE_+$ and let $A :=
\Ga(\psi)B \in \cE_\cO$. Then
\[
    0 =
    \< \hat C, \hat A\>_{\cH}
    =
    \< \hat C, \Pi(\Ga(\psi) B) \>_{\cH}
    =
    \< \Th C, \Ga(\psi) B \>_{\cE} \, .
\]
Since $\Ga(\psi)^{-1}$ is unitary on $\cE$, apply it to the inner
product to yield
\[
    \< \Ga(\psi^{-1}) \Th C, B\>_{\cE} = 0
    \quad (\forall \, B \in \cE_+).
\]
Therefore $\Ga(\psi^{-1}) \Th C$ is orthogonal (in $\cE$) to the
entire subspace $\cE_+$.

First, suppose that $[\Ga(\psi^{-1}), \Th] = 0$. Then we infer
\[
    0 =
    \< \Th \Ga(\psi^{-1}) C, B \>_{\cE}
    =
    \< \hat \Ga(\psi^{-1}) \hat C, \hat B\>_{\cH}
    \quad
    (\forall \, \hat B \in \Pi(E_+)),
\]
i.e. $\hat C \in \ker \hat \Ga(\psi^{-1})$. Therefore,
\bel{kernelGamma_psi}
    \lrp{ \Pi(\cE_\cO) }^\perp = \ker \hat \Ga(\psi^{-1}) \, .
\ee
Since $[\Ga(\psi^{-1}), \Th] = 0$ then Theorem \ref{thm:unit}
implies that $\hat \Ga(\psi)$ is unitary, hence the kernel of $\hat
\Ga(\psi^{-1})$ is trivial and $\Pi(\cE_\cO)$ is dense in $\cH$. We
have thus completed the proof in this case.

Now, assume that $\Ga(\psi) \Th = \Th \Ga(\psi^{-1})$. Example
\ref{ex:reflectediso} implies that $\hat \Ga(\psi)$ exists and is
self-adjoint on $\cH$, and moreover (by the same argument used
above),
\[
    \lrp{ \Pi(\cE_\cO) }^\perp = \ker \hat \Ga(\psi) \, .
\]
If $\psi = \psi_t$ where $\{ \psi_s \}$ is a one-parameter group of
isometries, and if $\hat \Ga(\psi_t)$ is a strongly continuous
semigroup then by Stone's theorem, $\hat \Ga(\psi_t) = e^{-t K}$ for
$K$ self-adjoint. Since $e^{-t K}$ clearly has zero kernel, the
proof is also complete in the second case.
\end{proof}

\begin{corollary}
The set $\cO_{+,T}$ is a quantization domain.
\end{corollary}

The problem of characterizing all quantization domains appears to be open.

%%%%%%%%%%%%%%%%%%%%%%%%%%%%%%%%%%%%%%%%%%%%%%%%%%%%%%%%%%%%
\subsection{Construction of the Hamiltonian and ground state}
\label{sec:HamiltonianAndGS}
%%%%%%%%%%%%%%%%%%%%%%%%%%%%%%%%%%%%%%%%%%%%%%%%%%%%%%%%%%%%

\begin{theorem}[Time-translation semigroup] \label{thm:kernel}
Let $\xi = \d / \d t$ be the time-translation Killing field on the
static spacetime $M$. Let the associated one-parameter group of
isometries be denoted $\phi_t : M \to M$. For $t \geq 0$, $U(t) =
\Gamma(\phi_t)$ has a quantization, which we denote $R(t)$. Further,
$R(t)$ is a well-defined one-parameter family of self-adjoint
operators on $\H$ satisfying the semigroup law.
\end{theorem}

\begin{proof}
Lemma \ref{lem:isometry1} implies that $U(t)$ is unitary on $\cE$,
and it is clearly a one-parameter group. Also,
\[
    \phi_t \circ \th = \th \circ \phi_{-t}
\]
and $U(t) \cE_+ \subset \cE_+$ for $t \geq 0$, so this is a
\emph{reflected isometry}; see Example \ref{ex:reflectediso}.
Theorem \ref{thm:sa} implies $R(t) = \hat U(t)$ is a self-adjoint
transformation on $\cH$ for $t \geq 0$, which satisfies the group
law
\[
    R(t) R(s) = R(t+s) \ \te{ for }\  t,s \geq 0
\]
wherever it is defined.
\end{proof}

\begin{theorem}[Hamiltonian and ground state] \label{thm:sg}
$R(t)$ is a strongly continuous contraction semigroup, which leaves invariant
the vector $\Omega_0 = \hat 1$. There exists a densely defined, positive,
self-adjoint operator $H$ such that
\[
    R(t) = \exp(-tH), \ \te{ and }\ H \Om_0 = 0.
\]
Thus $\Om_0$ is a quantum-mechanical ground state.
\end{theorem}

\begin{proof}
It is immediate that $R(t)\Om_0 = \Om_0$. The contraction property 
$R(t) \leq I$ follows from the multiple reflection method, as explained in \cite{GJ}. The remaining statements are consequences of Stone's theorem. 
\end{proof}

The operator $H$ is the quantum mechanical generator (in the
Euclidean picture) of translations in the direction $\xi$. When $\xi = \d / \d t$, then $H$ is called the \emph{Hamiltonian}.

%%%%%%%%%%%%%%%%%%%%%%%%%%%%%%%%%%%%%%%%%%%%%%%%%%%%%%%%%%%%
\subsection{Feynman-Kac theorem}
%%%%%%%%%%%%%%%%%%%%%%%%%%%%%%%%%%%%%%%%%%%%%%%%%%%%%%%%%%%%

\begin{theorem}[Feynman-Kac] \label{thm:fk}
Let $\hat A, \hat B \in \H$, and let $H$ be the Hamiltonian
constructed in Theorem \ref{thm:sg}. Each matrix element of the heat
kernel $e^{-t H}$ is given by a Euclidean functional integral,
\bel{fkformula}
    \< \hat A, e^{-t H} \hat B\>_{\H}
    =
    \int \overline{\Theta A}\, U(t) B \ d\mu(\Phi) \, .
\ee
\end{theorem}

The right-hand side of \er{fkformula} is the Euclidean path integral
\cite{Feynman} of quantum field theory. Mark Kac' method
\cite{Kac,KacWiener} for calculating the distribution of the
integral $\int_0^T v(X_t) dt$, where $v$ is a function defined on
the state space of a Markov process $X$, gives a rigorous version of Feynman's
work, valid at imaginary time.

In the present setup, \er{fkformula} requires no proof, since the
functional integral on the right-hand side is how we defined the
matrix element on the left-hand side. However, some work is required
(even for flat spacetime, $M = \R^d$) to see that the Hilbert space
and Hamiltonian given by this procedure take the usual form arising
in physics. This is true, and was carried out for $\R^d$ by
Osterwalder and Schrader \cite{OS} and summarized in
\cite[Ch.~6]{GJ}.

Since $H$ is positive and self-adjoint, the heat kernels can be
analytically continued $t\to it$. We therefore define the
\emph{Schr\"odinger group} acting on $\cH$ to be the unitary group
\[
        R(it)=e^{-itH}\;.
\]
Given a time-zero field operator, action of the Schr\"odinger group
then defines the corresponding real-time field.

For flat spacetimes in $d \leq 3$ it is known \citep{GJ} that
Theorem \ref{thm:fk} has a generalization to non-Gaussian integrals,
i.e. interacting quantum field theories:
\begin{align} \lab{fk}
    \< \hat A, e^{-t H_V} \hat B\>_{\H}
    &=
    \Big\< \Th A, \exp\lrpBig{-\int_0^t dt' \int d{\vec x}\ V(\Phi(\vec x, t'))}
B_t \Big\>_{\cE}
    \cr &=
    \int \overline{\Th A}\, e^{-S_{0,t}^V}\, B_t\ d\mu(\Phi)\, .
\end{align}
Construction the non-Gaussian measure \er{fk} in finite volume
can presumably be completed by a straightforward extension of present methods,
while the infinite-volume limit seems to require a cluster expansion. Work is in
progress to address these issues for curved spacetimes.

%%%%%%%%%%%%%%%%%%%%%%%%%%%%%%%%%%%%%%%%%%%%%%%%%%%%%%%%%%%%
\subsection{Quantization of subgroups of the isometry group}
%%%%%%%%%%%%%%%%%%%%%%%%%%%%%%%%%%%%%%%%%%%%%%%%%%%%%%%%%%%%
\label{sec:SubgroupsOfIsometries}

Physics dictates that after
quantization, a spacetime symmetry with $p$ parameters should correspond to a
unitary representation of a $p$-dimensional Lie group acting on $\H$. The group of spacetime symmetries for Euclidean quantum field theory should be related to the group for the real-time theory by analytic continuation; this was shown for flat spacetime
by Klein and Landau \cite{KleinLandau}. For curved spacetimes, 
no such construction is known, and due to the intrinsic interest of such a construction, we give further details, and show that the
methods already discussed in this paper suffice to give a
unitary representation of the purely spatial symmetries on $\cH$.

Example \ref{ex:reflection-invariant} introduced reflection-invariant isometries.
We now discuss an important subclass of these, the purely spatial isometries,
which are guaranteed to have well-defined quantizations.
We continue to assume we have a static manifold $M$ with notation as
in Example \ref{ex-theta-a}.
There is a natural subgroup $\Gspace$ of $G = \Iso(M)$
consisting of isometries which map each spatial section
into itself. We term these {\bf purely spatial} isometries.
The classic constructions \cite{JLW} of finite-volume interactions
in two dimensions work on a cylinder $M = S^1 \times \R$,
in which case $\Gspace$ is the subgroup of $\Iso(S^1\times \R)$
corresponding to rotations around the central axis.

Define the centralizer of $\th$ in $G$ as usual:
\[
    C_G(\th) \equiv
    \{ \phi \in G \mid \phi \th = \th \phi \} \, .
\]
Note that
\bel{Gspace}
    \Gspace \subset \Iso(M,\Om_+) \cap C_G(\th),
\ee
and although the right hand side of \er{Gspace}
is not a subgroup, this is a compact way
of expressing that the elements are positive-time invariant and
null-invariant.
Since $\Gspace \subset G$ as a Lie subgroup, $\gsp =
\Lie(\Gspace)$ is a subalgebra of $\fK$, the Lie algebra
of global Killing fields.

Consider the restriction of the unitary representation
$\Ga$ to the subgroup $\Gspace$. By a standard construction, the
derivative $D\Gamma$ is a unitary Lie algebra representation of $\gsp$ on $\cE$,
for which $\cE_+$ is an invariant subspace. The latter property is
crucial; if $\cE_+$ is not an invariant subspace for an operator, then
that operator does not have a quantization.

As with many aspects of Osterwalder-Schrader quantization, a commutative
diagram is helpful:
\bel{Uu-diagram}
    \xymatrix{
    \Gspace \ar[r]^{\Ga} \ar[d]_{\Lie} &  \cU(\cE) \ar[d]^{\Lie}  \\
    \gsp \ar[r]_{D\Gamma} & \mathfrak{u}(\cE)
    }
\ee
Note that $\cU(\cE)$ is an infinite-dimensional Lie group. Further, there are
delicate analytic questions involving the domains of the symmetric operators
in $\mathfrak{u}(\cE)$. In the present paper we investigate only the algebraic
structure.

By Theorem \ref{thm:unit}, each one-parameter unitary group $U(t)$
on $\cE_+$ coming from a one-parameter subgroup of $\Gspace$ has a
well-defined quantization $\hat U(t)$ which is a unitary group on
$\cH$. The methods of Section \ref{sec:strongcont} establish strong
continuity for these unitary groups, so their generators are densely-defined
self-adjoint operators as guaranteed by Stone's theorem.

Suppose that $[X,Y] = Z$ for three elements $X,Y,Z \in
\gsp$. Let $\hat X : \cH \to \cH$ be the quantization of $D\Ga(X)$,
and similarly for $Y$ and $Z$.  Our assumptions guarantee that
$[D\Ga(X), D\Ga(Y)] = D\Ga(Z)$ is null-invariant, therefore we have
\bel{LieAlg-hatX}
    [\hat X, \hat Y] = \hat Z ,
\ee
valid on the domain of vectors in $\cH$ where the expressions are defined.

One-parameter subgroups coming from $\Gspace$ always admit unitary
representations on $\cH$, but for other subgroups of $G$, the analogous theory
is much more subtle.
Since any element of $\fK$ is a
vector field acting on functions as a differential operator, it is local
(does not change supports) and hence positive-time invariant, so
quantization applied directly to infinitesimal generators may be possible.
There, one runs into delicate domain issues. A discussion of the domains of some
self-adjoint operators obtained by this procedure was given in Section
\ref{sec:quantizationdomains}, and some variant of this could possibly be used
to treat the domains of the quantized generators.

When applied to isometry groups, Osterwalder-Schrader
quantization of operators involves the
procedure of taking the derivative of a representation, applied to the
infinite-dimensional group $\cU(\cE)$. Thus, it is not surprising that it is
functorial, adding to its intrinsic mathematical interest. These connections are
likely to lead to an interesting new direction in representation theory,
especially for noncompact groups.

%%%%%%%%%%%%%%%%%%%%%%%%%%%%%%%%%%%%%%%%%%%%%%%%%%%%%%%%%%%%
\section{Variation of the Metric} \label{sec:VariationOfMetric}
%%%%%%%%%%%%%%%%%%%%%%%%%%%%%%%%%%%%%%%%%%%%%%%%%%%%%%%%%%%%
\subsection{Metric dependence of matrix elements in quantum field theory}
%%%%%%%%%%%%%%%%%%%%%%%%%%%%%%%%%%%%%%%%%%%%%%%%%%%%%%%%%%%%

We wish to obtain rigorous analytic control over how
quantum field theory on a curved background depends upon the metric.

\begin{definition}[Stable family]
Let $M_\la$ denote the Riemannian manifold diffeomorphic to $\R \times S$,
endowed with the product metric
\bel{metsigma}
	ds^2_\la := dt^2 + G_{\m\n}(\la) dx^\m dx^\n,
\ee
where $G(\la)$ is a metric on $S$, and $G_{\m\n}(\la)$ depends smoothly on $\la \in \R$. We refer to a family
$\{ M_\la \}_{\la \in \R}$ satisfying these properties as a \emph{stable
family}. We denote the full metric \er{metsigma} as $g(\la)$ or $g_\la$.
\end{definition}

For a stable family, it is clearly possible
to choose $\Om_\pm, \Sigma$ in a way that is independent
of $\la$. Let $t$ denote the coordinate which is defined
so that $t|_\Sigma = 0$ and $\xi = \d / \d t$. Then the data $(\Om_-, \Sigma,
\Om_+, \xi, t)$ is constant in $\la$.

However, the Hilbert spaces $L^2(M_\la)$, the covariance
\[
	C(\la) = C_\la := \lrp{-\De_{g(\la)} + m^2}^{-1} \, ,
\]
and the test function space $H_{-1}(M_\la)$ all depend upon $\la$, as does the
Gaussian measure described in section \ref{sec:fundamentals}. These dependences
create many subtleties in the quantization procedure. In particular, the usual
theory of smooth or analytic families of bounded operators does not apply to the
family of operators $\la \to C(\la)$, because if $\la \ne \la'$ then $C(\la)$
and $C(\la')$ act on different Hilbert spaces. It is clearly of interest to have
some framework in which we can make sense out of the statement ``$\la \to
C(\la)$ is smooth.'' More generally, we would like a framework to analyze the
$\la$-dependence of the Osterwalder-Schrader quantization.

Our approach to this set of problems is based on the observation that, for a
stable family, there exist test functions $f : M \to \R$ which are elements of
$H_{-1}(M_\la)$ for all $\la$. For example, 
\bel{H-1}
	C_c^\infty(M)
	\ \subset\
	\cH_{-1}
	:=
	\bigcap_{\la \in \R} H_{-1}(M_\la) \, .
\ee
Such test functions can be used to give meaning to formally ill-defined expressions such as $\d C_\la / \d \la$. To give meaning to the naive expression
\bel{cprime}
	\f{\d C_\la}{\d \la} \, f
	\ :=\
	\lim_{\eps \to 0} \f{1}{\eps} \lrp{ C_\la f - C_{\la + \eps} f } ,
\ee
we must specify the topology in which the limit is to be taken. Suppose that $f
\in C_c^\infty$ as before. A natural choice is the topology of $L^2(M_\la)$, but
some justification is necessary in the noncompact case. Clearly $C_\la f \in
L^2(M_\la)$, but it is not clear that $C_{\la + \eps} f$ also determines an
element of $L^2(M_\la)$. After all, the covariance operators are nonlocal, and
$C_{\la + \eps} f$ generally does not have compact support (unless of course $M$
itself is compact).

In order that the limit \er{cprime} can be taken in the topology of
$L^2(M_\la)$, it is necessary and sufficient that $\exists\,\eps_1 > 0$ such
that $C_{\la +\eps} f \in L^2(M_\la)$ for all $\eps < \eps_1$. In other words,
the limit \er{cprime} makes sense iff $F(\eps) \equiv \int_M |C_{\la + \eps}f|^2
\sqrt{|g_\la|} \, dx < \infty$ for all $\eps <\eps_1$. Since obviously $F(0)
<\infty$, it suffices to show $F(\eps)$ is continuous at $\eps = 0$. If we write
the expressions in terms of coordinate charts and assume $f > 0$, then we can
translate the problem into one of classical analysis. Indeed,
\bel{Feps}
	F(\eps)= \int_M dx \sqrt{|g_\la(x)|}
	\lrp{ \int_{\supp f} dy  \sqrt{|g_{\la+\eps}(y)|} \, C_{\la + \eps}(x,y) f(y) }^2 .
\ee
Thus the condition for differentiability of $F(\eps)$ at $\eps = 0$ becomes one
of ``differentiating under the integral,'' which can be treated by standard
methods. The overall conclusion: if $F(\eps)$ is continuous at $\eps = 0$, then
\er{cprime} makes sense. Anticipating what is to come, this condition implies
that \er{9.1.33} also makes sense.

We now return to the study of the full quantum theory on $M_\la$. Define
\[
	\cE_\la := L^2(d\mu_\la)
\]
where $d\mu_\la$ is the unique Gaussian probability measure associated to
$C(\la)$ by Minlos' theorem.\footnote{As before,
$
    E_{+,\la} = \mathrm{span} \big\{ e^{i \Phi(f)} \, \mid \,
    f \in H_{-1}(M_\la),
	\ \supp(f) \subset \Om_+ \big\},
$
with completion
\[
	\cE_{+,\la} = \overline{E_{+,\la}} \, .
\]
Also define $E_\la$ to be the (incomplete) linear span of $e^{i \Phi(f)}$ for
$f \in H_{-1}(M_\la)$.}
If $f\in \cH_{-1}$, then
\bel{CanonicalElement}
    A_{f,\la} = \no{e^{-i \Phi(f)}}_{C(\la)}
\ee
defines a canonical element of $\cE_\la$ for each $\la$.
Then
\bel{FieldTheoryMatrixElement}
	\< A_{f,\la}, A_{g,\la}\>_{\cE, \la}
	=
	\exp\lrp{ \left\<f, C_\la g\right\>_{L^2(M_\la)} }.
\ee

\begin{lemma}[Smoothness of covariance]
Assume that $\{ M_\la \}_{\la \in \R}$ is a stable family. Then
$\< f, C(\la) g\>_{L^2(M_\la)}$ is a smooth function of $\la$, for any $f, g
\in C_c^\infty(M)$.
\end{lemma}

\begin{proof}
The integral $\< f, C(\la) g\>_{L^2(M_\la)} = \int_M  \ol{f} C_\la g \,
\sqrt{|g_\la|}\, dx$ is localized over the support of $f$, which is compact. The
dominated convergence theorem shows that we can interchange $\d / \d \la$ with
the integral.
\end{proof}

It follows immediately that the matrix element \er{FieldTheoryMatrixElement} on
$\cE$ of the canonical elements  $A_{f,\la}$ and $A_{g,\la}$ is a smooth
function of the parameter $\la$.

When we change $\la$, the measure $d\mu_\la$ follows a path in the space of all
Gaussian measures. This change in the measure can be controlled through operator
estimates on the covariance. Using  formula 9.1.33 from \citep[p.~208]{GJ} we
have:
\bel{9.1.33}
    \f{d}{d\la} \int A \ d\mu_\la =
    \f12 \int (\De_{d C/d\la} A) \ d\mu_\la \, .
\ee
In particular, if $C(\la)$ is smooth then so is $\int A \ d\phi_{C(\la)}$. Here
we must interpret $d C/d\la$ as in the discussion following \er{cprime}.

The null space $\cN_\la$ of OS quantization also depends on the metric, as we
discuss presently. When it is necessary to distinguish the time direction, we
denote local coordinates by $x = (\vec x, t)$. The subspace of $\cN_\la$
corresponding to monomials in the field is canonically isomorphic to the space
of test functions $f$ such that\footnote{For integrals such as this one, we can
factorize the Laplacian as in Sec.~\ref{sec:localization}.}
\bel{kernelf}
    \int_M f(\vec x,-t) \lrp{ C_\la \,f }(\vec x,t)
    \sqrt{ |g_\la(\vec x)| } \, d x
    = 0 \, .
\ee
All of the quantities in the integrand \er{kernelf}
which depend on $\la$ do so smoothly. Assuming the applicability of dominated
convergence arguments similar to those used above, it should be possible to show
that $\cN_\la$ varies continuously in the Hilbert Grassmannian, but we do not
address this here.

For each $\la$, the Osterwalder-Schrader theory gives unambiguously a quantization
\[
	\cH(\la) \ \equiv \  \overline{\cE_{+, \la} / \cN_\la } .
\]

\begin{theorem}[Smoothness of matrix elements in $\cH$]\label{thm:smoothness}
Assume that $\{ M_\la \}_{\la \in \R}$ is a stable family.
Define the canonical element $A_{f,\la}$ as in \er{CanonicalElement}. Then
\[
		\la \to
		\< {\hat A_{f,\la}}
		,\,
		R_\la(t) {\hat A_{g,\la}}
		\>_{\cH(\la)}
\]
is smooth.
\end{theorem}

\begin{proof} Calculate $\< \hat A, R_\la(t) \hat B\>_{\cH(\la)} =
\exp{\< \th f, (C_\la h) \circ \phi_{\la,t}^{-1}\>}$,
where $\phi_{\la,t}$ is the time $t$ map of the Killing field $\d / \d t$ on the
spacetime $M_\la$. Since $f$ has compact support, the dominated convergence
theorem applies to the integral $\< \th f, (C_\la h) \circ \phi_{\la,t}^{-1}\>$.
\end{proof}

One class of examples which merits further consideration is the class
formulated on $M = \R^{d+1}$ with $ds^2 = dt^2 + g(\la)_{ij} dx^i dx^j,$ $i,j =
1 \ldots d$. Assume that $G(\la)_{ij}$ depends analytically on $\la \in \C$, and
to order zero it is the flat metric $\de_{ij}$. Theorem \ref{thm:smoothness}
implies that the matrix elements of $H$ have a well-defined series expansion
about $\la = 0$, and we know that precisely at $\la = 0$ they take their usual
flat-space values.

%%%%%%%%%%%%%%%%%%%%%%%%%%%%%%%%%%%%%%%%%%%%%%%%%%%%%%%%%%%%
\subsection{Stably symmetric variations} \label{sec:preserve-NumKF}
%%%%%%%%%%%%%%%%%%%%%%%%%%%%%%%%%%%%%%%%%%%%%%%%%%%%%%%%%%%%

It is of interest to extend the considerations of the previous section to the
quantizations of symmetry generators. For this we continue to consider
variations of an ultrastatic metric, as in equation \er{metsigma}. One important
aspect of the quantization that is generally \emph{not} $\la$-invariant is the
symmetry structure of the Riemannian manifold. We assume $M = \R \times M'$,
where $M'$ is a Riemannian manifold with metric $g_{\m\n}(\la)$. In this section
we study a special case in which the perturbation does not break the symmetry.
Let $\fK_\la$ denote the algebra of global Killing fields on $(M', g(\la))$. In
certain very special cases we may have the following.

\begin{definition}[Stable symmetry]\label{def:stablesymmetry}
The family of metrics $\la \to g(\la)$ is said to be {\bf stably
symmetric} over the subinterval $I \subset \R$ if for each $\la \in I$, there
exists a basis $\{ \xi_i(\la) : 1 \leq i \leq n \}$ of $\fK_\la$, and the family
of bases can be chosen in such a way that  $\la \to \xi_i(\la)$ is smooth
$\forall \ i$.
\end{definition}

Equivalently, the condition of stable symmetry is that
$\fK_\la = KF(M_\la)$ gives
a rank $n$ vector bundle over $\R$ (or some subinterval thereof)
and we have chosen a complete set $\{\xi_i : i = 1 \ldots n \}$
of smooth sections.

\begin{example}[curvature variation]
\label{example:cchm}
The most general constant-curvature hyperbolic metric on $\bbH$
has arc length
\bel{schypm}
    ds = \f{c}{\Im(z)} |dz|
\ee
and curvature $-c^{-2}$. Consider the spacetime $\R \times
\bbH(c)$ where $\bbH(c)$ is the upper half-plane with metric
\er{schypm}. Variation of the curvature parameter $c$ satisfies the assumptions
of definition \ref{def:stablesymmetry}.
\end{example}

\begin{example}[ADM mass, charge, etc.]
Many spacetimes considered in physics seem to have the property of stable
symmetry under variation of parameters, at least for certain ranges of those
parameters. For the Euclidean continuation of the
Reissner-Nordstr\"{o}m black hole, where $\la$ plays the role of either mass $m$
or charge $e$, one may observe that the assumptions of definition
\ref{def:stablesymmetry} hold. However, the Euclidean RN metrics are not
ultrastatic as was assumed above. Therefore, it would be interesting to
extend the analysis of this section to static metrics of the form $F(\la,x) dt^2
+ G(\la,x) dx^2$, where $x$ is a $d-1$ dimensional coordinate.
\end{example}

For each $i, \la$, the Killing field $\xi_i(\la)$ gives rise to a
one-parameter group of isometries on $M$, which we denote by
$\phi_{i, \la, x} \in \Iso(M)$, where $x \in \R$ is the flow
parameter. These flows act on the spatial section of $M$
for each fixed time; they are \emph{purely spatial isometries} in
the sense considered above. Therefore, the
map
\bel{Tmap}
    T_i(\la,x) = \Ga(\phi_{i, \la, x}) : \cE \To \cE \, .
\ee
is positive-time invariant, null-invariant, and has a unitary quantization
\bel{quant-Ti}
    {\hat T}_i(\la,x) : \H \To \H \,.
\ee

None of the following constructions depend on $i$, so for the
moment we fix $i$ and suppress it in the notation.
Since each $T(\la,x)$ depends on a Killing field $\xi$, the first
step is to determine how the Killing fields vary as a function of
the metric. Since the Killing fields are solutions to a
first-order partial differential equation, one possible method of
attack could proceed by exploiting known regularity properties of
solutions to that equation. If one were to pursue that, some
simplification may be possible due to the fact that a Killing
field is completely determined by its first-order data at a point.
We obtain a more direct proof.

The $T$ operators depend on the Killing field through its
associated one-parameter flow. For each fixed $\la$, the construction
gives a one-parameter subgroup (in particular, a curve)
in $\Gspace$. If we vary $\la \in [a,b]$, we have a free homotopy
between two paths in $\Gspace$. Each cross-section of this homotopy,
such as $\la \to \phi_{\la,x}(p)$ with the pair $(x,p)$ held
fixed, describes a continuous path in a particular spatial
section of $M$.

\begin{theorem}
Assume stable symmetry and define $T(\la,x)$ as in \er{Tmap}. Then for each $x$
(held fixed), the map
\[
    \la \Mapsto {\hat T}(\la,x) \in \cU(\cH)
\]
is a strongly continuous operator-valued function of $\la$.
\end{theorem}

\begin{proof}
First, we claim that $\la \to \phi_{\la,x}$ is continuous in the
compact-open topology. The latter follows from standard regularity
theorems for solutions of ODEs, since we have assumed $\la \to
\xi(\la)$ is smooth, and $\phi_{\la,x}(p)$ is the solution curve
of the differential operator $\xi(\la)_p$. Theorem
\ref{thm:strongcont} implies that $\Ga(\phi_{\la,x}) \in \cU(\cE)$
is strongly continuous with respect
to $\la$. By theorem \ref{prop:contraction}, the embedding of
bounded operators on $\cE$ into $\cB(\cH)$ is norm-continuous. Composing these
continuous maps gives the desired result.
\end{proof}

%%%%%%%%%%%%%%%%%%%%%%%%%%%%%%%%%%%%%%%%%%%%%%%%%%%%%%%%%%%%
\section{Sharp-time Localization} \label{sec:localization}
%%%%%%%%%%%%%%%%%%%%%%%%%%%%%%%%%%%%%%%%%%%%%%%%%%%%%%%%%%%%

The goal of this section is to establish an analog of \citep[Theorem 6.2.6]{GJ} for quantization in curved space, and to show that the
Hilbert space of Euclidean quantum field theory may be expressed in terms of
data local to the zero-time slice. This is known as \emph{sharp-time localization}. We first define the type of spacetime to
which our results apply.

\begin{definition}\label{def:localization}
A {\bf quantizable static spacetime} is a complete, connected
Riemannian manifold $M$ with a globally defined (smooth) Killing field $\xi$ which is orthogonal to a codimension one
hypersurface $\Sigma \subset M$, such that the orbits of $\xi$ are
complete and each orbit intersects $\Sigma$ exactly once.
\end{definition}

Under the assumptions for a quantizable static spacetime, but with
Lorentz signature, Ishibashi and Wald \cite{Ishi-Wald} have shown
that the Klein-Gordon equation gives sensible classical dynamics,
for sufficiently nice initial data. These assumptions guarantee that we are in the situation of Definition \ref{time-reflection}.

The main difficulty in establishing sharp-time localization comes when trying to prove the analog of
formula (6.2.16) of \cite{GJ} in the curved space case, which would imply that
the restriction to $\cE_0$ of the quantization map is surjective. The proof
given in \cite{GJ} relies on the formula (6.2.15) from Prop.~6.2.5, and it is
the latter formula that we must generalize.

%%%%%%%%%%%%%%%%%%%%%%%%%%%%%%%%%%%%%%%%%%%%%%%%%%%%%%%%%%%%
\subsection{Localization on flat spacetime}
%%%%%%%%%%%%%%%%%%%%%%%%%%%%%%%%%%%%%%%%%%%%%%%%%%%%%%%%%%%%

The Euclidean propagator on
$\R^d$ is given explicitly by the momentum representation
\[
        C(x;y) = C(x-y) =
        \frac{1}{\lrp{2\pi}^d}
        \int_{\R^d} \frac{1}{p^2+m^2}\; e^{-ip\cdot (x-y)} \; dp\;,
\]
for $x,p\in\R^d$. Let $f = f(\vec x)$ denote a function on
$\R^{d-1}$, and define
\[
    f_t(\vec x, t') = f(\vec x) \de(t - t') \, .
\]

\begin{theorem}[Flat-space localization]\label{theorem:GJ-6.2.15}
Let $M = \R^d$ with the standard Euclidean metric. Then
\[
    \< f_t, C g_s \>_{L^2(\R^d)}
    =
    \Big\< f, \f{1}{2\mu}\, e^{(t-s) \mu} g \Big\>_{L^2(\R^{d-1})}
\]
where $\mu$ is the operator with momentum-space kernel $\mu(\vec p) =
\lrp{\vpsqr + m^2}^{1/2}$.
\end{theorem}

%%%%%%%%%%%%%%%%%%%%%%%%%%%%%%%%%%%%%%%%%%%%%%%%%%%%%%%%%%%%
\subsection{Splitting the Laplacian on static spacetimes}
%%%%%%%%%%%%%%%%%%%%%%%%%%%%%%%%%%%%%%%%%%%%%%%%%%%%%%%%%%%%
Consider a quantizable static space-time $M$, defined in Definition
\ref{def:localization}. Use Latin indices $a,b,$ etc. to run from $0$ to $d-1$
and Greek indices $\mu,\nu = 1 \ldots d-1$. Denote the spatial
coordinates by
\[
    \vx = (x^1,\ldots, x^{d-1}) = (x^\mu)\;,
\]
and set $t =x^0$. Write $g$ in manifestly static form,
\bel{eq:manifestlystatic}
    g_{ab} =
    \begin{pmatrix}
    F & 0 \\
    0 & G_{\m\n}
    \end{pmatrix} \ ,
    \ \
    \text{ with inverse }
    \ \
    g^{ab} =
    \begin{pmatrix}
    1/F & 0 \\
    0 & G^{\m\n}
    \end{pmatrix} \ .
\ee
where $F$ and $G$ depend only on $\vec x$, and not on $t = x^0$.
It is then clear that
\bel{eq:scrg}
    \scrg := \det(g_{ab}) = FG, \te{ where }
    G = \det(G_{\m\n}) \, .
\ee
It follows that $g^{0\n} = g^{\m 0} = 0$, and $g^{00} = {F}^{-1}
={g_{00}}^{-1}$, does not depend upon time. Using the formula,
$\Delta f = \scrg^{-1/2} \d_a\lrp{\scrg^{1/2}\, g^{ab} \d_b f}$, the
Laplacian on $M$ may be seen to be
\begin{eqnarray}
    \label{CoordinateLaplacian1}
    \De_M &=&
    \f{1}{F} \d_t^2 + Q,
    \ \ \te{ where }
    \\
    Q &:=&
    \f{1}{\sqrt{\scrg}} \d_\m \lrp{ \sqrt{\scrg} \, G^{\m\n} \, \d_\n }
    \, .
    \label{StaticForm}
\end{eqnarray}
The operator $Q$ is related to the
Laplacian $\Delta_\Sigma$ for the induced
metric on $\Sigma$. Applying the product rule to \er{CoordinateLaplacian1}
yields
\bel{easy-Q}
    Q = \f{1}{2} \d_\alpha (\ln F) \, G^{\alpha\beta} \d_\beta + \Delta_\Sigma \, .
\ee
Note that a formula generalizing \er{easy-Q} to ``warped products'' appears in
Bertola et.al. \cite{Bertola:2000}.

In order that the operator $\mu = (-Q + m^2)^{1/2}$ exists for all $m^2 > 0$, we require that $-Q$ is a positive,
self-adjoint operator on an appropriately-defined Hilbert space. The correct
Hilbert space is
\bel{eq:K_Sigma}
        \cK_\Sigma \ := \ L^2(\Sigma, \sqrt{\scrg} \, dx)\;.
\ee
Here $\sqrt{\scrg} \, dx$ denotes the Borel measure on $\Sigma$ which has
the indicated form in each local coordinate system, and $\scrg = FG$ as in
eq.~\er{eq:scrg}.

Spectral theory of the operator $-Q$ considered on $\cK_\Sigma$ is
mathematically equivalent to that of the ``wave operator'' $A$ defined by Wald
\cite{Wald79,Wald:1980} and Wald and Ishibashi \cite{Ishi-Wald}. In those
references, the Klein-Gordon equation has the form $(\d_t^2 + A)\phi = 0$. The
relation between Wald's notation and ours is that $Q = -(1/F) A - m^2$, and
Wald's function $V$ is our $F^{1/2}$.  As pointed out by Wald, we have the
following,

\begin{theorem}[\bf $Q$ is symmetric and negative] \label{thm:Qsymmetric}
Let $(M, g_{ab})$ be a quantizable static spacetime. Then $-Q$ is a symmetric,
positive operator on the domain $C_c^\infty(\Sigma)\subset \cK_\Sigma$.
\end{theorem}

\begin{proof}
It is easy to see that $Q$ is symmetric on $C_c^\infty(\Sigma)$
with the metric of $\cK_\Sigma$; it remains to show $-Q \geq 0$ on
the same domain.
Using \er{StaticForm}, the associated quadratic form is
\begin{eqnarray*}
    \< f, (-Q)f\>_{\cK_\Sigma} &=&
    - \int \overline{f} \f{1}{\sqrt{\scrg}} \d_\mu \lrp{ \sqrt{\scrg} \,
G^{\mu\nu} \d_\nu f }
    \ \sqrt{\scrg} \, dx
    \\
    &=& \int {\left\| \nabla f \right\|}_G^2  \ \sqrt{\scrg} \, dx  \geq 0 \,
.
\end{eqnarray*}
where we used integration by parts to go from the first line to the second.
\end{proof}

\subsection{Hyperbolic space}  \label{sec:hyperbolic-space}

It is instructive to calculate $Q$ in the explicit example of
$\hyp{d}$, often called \emph{Euclidean AdS} in the physics literature because
its analytic continuation is the Anti-de Sitter spacetime. The metric is
\[
	ds^2 = r^{-2} \sum_{i=0}^{d-1} dx_i^2, \quad
	r = x_{d-1} \, .
\]
The hyperbolic Laplacian in $d$ dimensions
is (see for instance \cite{Beardon}):
\bel{eq:hyp-lapl}
    \De_{\hyp{d}} = (2-d) r \f{\d}{\d r}
    +
    r^2 \De_{\R^d} \, .
\ee
Any vector field $\d / \d x_i$ where $i \ne d-1$
is a static Killing field. We have set up
the coordinates so that it is convenient to define $t = x_0$ as
before, and we can quantize in the $t$ direction.

Comparing \er{StaticForm} with \er{eq:hyp-lapl}, we find that $F = r^{-2}$ and
\bel{split-Q-hyp}
    Q = (2-d) r \f{\d}{\d r}
    +
    r^2 \sum_{i=1}^{d-1} \f{\d^2}{\d x_i^2}
    = - r \f{\d}{\d r} + \De_{\hyp{d-1}} \, ,
\ee
which matches \er{easy-Q} perfectly.
We return to this example spacetime in Appendix \ref{sec:AdSGreensFunctions},
where we calculate its
Green function, and discuss the analytic continuation.

%%%%%%%%%%%%%%%%%%%%%%%%%%%%%%%%%%%%%%%%%%%%%%%%%%%%%%%%%%%%
\subsection{Curved space localization}\label{sec:curved-local}
%%%%%%%%%%%%%%%%%%%%%%%%%%%%%%%%%%%%%%%%%%%%%%%%%%%%%%%%%%%%

To generalize Theorem \ref{theorem:GJ-6.2.15} to curved space,
choose static coordinates $\vec x, t$ near the time-zero slice
$\Sigma$. If $f = f(\vec x)$ is a function on the slice $\Sigma$, we define
\[
    f_t(\vec x, t') = f(\vec x) \de(t - t'),
\]
which is a distribution on the patch of $M$ covered by this coordinate
chart. For the moment, we assume that this coordinate patch is the
region of interest.
By equation \er{StaticForm}, we infer
that the integral kernel $\cC(x,y)$ of the operator $C = (-\Delta + m^2)^{-1}$
is time-translation invariant, so that we may write
\[
    \cC(x,y) = \cC(\vx,\vy, x_0 - y_0) \, .
\]
In order to apply spectral theory to $Q$, we choose a self-adjoint extension
of the symmetric operator constructed by theorem \ref{thm:Qsymmetric}.
For definiteness, we may choose the Friedrichs extension, but any ambiguity
inherent in the choice of a self-adjoint extension will not enter into the
following analysis. We denote the self-adjoint extension also by $Q$, which
is an unbounded operator on $\cK_\Sigma$.
The following is a generalization of Theorem \ref{theorem:GJ-6.2.15} to
curved space.

\begin{theorem}[Localization of sharp-time integrals]\label{thm:dimreduc}
Let $M$ be a quantizable static spacetime (definition
\ref{def:localization}). Then:
\bel{eq:dimreduc}
    \< f_t, C g_s \>_{M}
    =
    \lra{ f,  \
    \lrpBig{F^{1/2}  \f{e^{- |t-s| \omega} }{2\omega} F^{1/2}} g
}_{\cK_\Sigma}\;,
\ee
where $\mu = \lrp{-Q + m^2}^{1/2}$ and $\omega = \lrp{\sqrt{F} \mu^2
\sqrt{F}}^{1/2}$.
Hence $C$ is reflection positive on $L^2({M})$.
\end{theorem}

\begin{proof}
Because $M$ was assumed to be a quantizable static spacetime,
$F = \< \xi,\xi \>_\Sigma \geq 0$. Moreover, if $F(p) = 0$
then $\xi_p = 0$, for any $p \in \Sigma$. A non-trivial Killing
field cannot vanish on an open set, so the zero-set of $F$ has
measure zero in $\Sigma$. From this we infer that multiplication by
the function $F^{-1}$ defines a (possibly-unbounded) but
densely-defined self-adjoint multiplication operator on
$\cK_\Sigma$.

For simplicity of notation, assume $f$ is real-valued. Perform a
partial Fourier transform with respect to the time variable:
\bel{Res1}
    \< f_t, C g_s \>_{M}
    =
    \int
    f(\vx)
    \lrp{
        \f{1}{2\pi}\int dE
        \f{ e^{iE (t-s)} }{F^{-1} E^2 - Q + m^2} g
        }(\vx)
        \;
    \sqrt{\scrg} \;
    dx
    \, .
\ee
Define $\mu \ :=\  \lrp{-Q + m^2}^{1/2},$
where the square root is defined through the spectral
calculus on $\cK_\Sigma$.  As a consequence of theorem \ref{thm:Qsymmetric},
$\mu$ and $\omega$ are positive, self-adjoint operators on $\cK_\Sigma$.
The integrand of \er{Res1} contains the operator:
\[
    \f{e^{iE (t-s)}}{F^{-1} E^2 + \mu^2}
    \, =\,
    \f{e^{iE (t-s)}}{F^{-1/2} \lrp{E^2 + F^{1/2} \mu^2 F^{1/2}} F^{-1/2}}
    \, =\,
    F^{1/2} \f{e^{iE (t-s)}}{E^2 + \omega^2} F^{1/2} \, .
\]

We next establish that $\omega$ is invertible. Since $\mu^2 > \eps I$, where $\eps > 0$, we have 
\[
    \omega^2 = \sqrt{F} \mu^2 \sqrt{F} > \eps F
\]
and therefore,
\[
    \omega^{-2} < \lrp{\sqrt{F} \mu^2 \sqrt{F}}^{-1} < \f{1}{\eps F} \, .
\]
Since $1/F$ is a densely defined operator on $\cK_\Sigma$, it
follows that $\omega^2$ (hence $\omega$) is invertible.
For $\la>0$,
    \bel{Res2}
        \int \f{e^{i E \tau}}{E^2 + \la^2} \, dE
        =
        \f{\pi e^{- |\tau|  \la} }{\la} \, .
    \ee
Decompose the operator $\omega$ according to its spectral
resolution, with $\omega=\int \la\, dP_\la$ and $I=\int dP_\la$ the
corresponding resolution of the identity, and apply \er{Res2} in
this decomposition to conclude \bel{op-eq1}
    \int \f{e^{iE (t-s)}}{F^{-1} E^2 + \mu^2}  \ dE
    \, =\,
    F^{1/2} \f{\pi e^{- |t-s| \omega} }{\omega} F^{1/2}
\ee
Inserting \er{op-eq1} into \er{Res1} gives
\begin{eqnarray}
    \< f_t, C g_s \>_{M}
    &=&
    \int_\Sigma
    \lrp{
    F^{1/2}f}(\vec x) \lrp{\f{e^{- |t-s| \omega} }{2\omega} (F^{1/2} g)}(\vec x)
        \;
    \sqrt{\scrg} \;   d\vx
    \nonumber
    \\
    &=&
    \lra{  f,  \
    F^{1/2} \f{e^{- |t-s| \omega} }{2\omega} F^{1/2} g }_{\cK_\Sigma} \, ,
    \label{eq:new-proof-RP}
\end{eqnarray}
also demonstrating reflection positivity.
\end{proof}

The operator $\omega^2$ may be calculated explicitly if the metric is known,
and is generally not much more complicated than $Q$.
For example, using the conventions of sec.~\ref{sec:hyperbolic-space}, one
may calculate $\omega^2$ for $\hyp{d}$:
\[
    \omega^2 = - \sum_{i=1}^{d-1} \d_i^2 + d\, r^{-1} \d_r + (m^2-d) r^{-2} \,
.
\]
For $\hyp{\,2}$, the eigenvalue problem $\omega^2 f = \la f$ becomes
a second-order ODE which is equivalent to Bessel's equation. The two linearly
independent solutions are
\[
	r^{3/2} J_{\frac{1}{2} \sqrt{4 m^2+1}}\lrp{r\sqrt{\lambda }}
	\quad \text{ and } \quad
	r^{3/2} Y_{\frac{1}{2} \sqrt{4 m^2+1}}\lrp{r \sqrt{\lambda }}.
\]
The spectrum of $\omega^2$ on $\hyp{2}$ is then $[0, +\infty)$.

Given a function $f$ on $\Sigma$, we obtain a
distribution $f_t$ supported at time $t$ as follows:
\[
    f_t(x, t') = f(x) \delta(t - t') .
\]
It may appear that this is not well-defined because it depends on
a coordinate. However, given a static Killing
vector, the global time coordinate is fixed up to an overall shift
by a constant, which we have determined by the choice of
an orthogonal hypersurface where $t = 0$. Thus a pair
$(p,t)$ where $p \in \Sigma$ and $t \in \R$ uniquely specify a
point in $M$.

\begin{theorem}[Localization of $\cH$]\label{thm:626}
Let $M$ be a quantizable
static spacetime. Then the vectors $\exp(i \Phi(f_0))$ lie in $\cE_+$, and
quantization maps the span of these vectors isometrically onto $\cH$.
\end{theorem}

\begin{proof}
Since $\cE_+$ is the \emph{closure} of the
set $E_+$ of vectors $\exp(i \Phi(f))$ with $\supp(f) \subset
\Om_+$, it follows that any sequence in $\cE_+$ which converges in the topology
of $\cE$ has its limit in $\cE_+$. The $L^2$ norm in $\cE$,
\[
    \int \abs{ e^{i \Phi(f)} - e^{i \Phi(g)} }^2 d\mu_C(\Phi)
    =
    2(1 - e^{ - \f12 \norm{f-g}_{-1}}),
\]
is controlled in terms of the norm $\norm{ \ }_{-1}$ on Sobolev
space, which is the space of test functions. This
will give us the first part of the theorem.

If $t > 0$, then there exists a sequence of smooth test functions
$\{g_n\}$ with compact, positive-time support such that
\[
    \lim_{n \to \infty} g_n = f_t
\]
in the Sobolev topology, hence $\exp(i \Phi(f_t)) \in \cE_+$.
Define the \emph{time-$t$ subspace}
$\cE_t \subset \cE_+$ to be the subspace generated by vectors of the
form $\exp(i \Phi(f_t))$. By taking the $t \to 0$ limit, we see that
$\exp(i\Phi(f_0)) \in \cE_+$ and the first part is proved.

It is straightforward to see that the quantization map $\Pi(A) \equiv
\hat{A}$ is isometric when restricted to vectors of the form
$\exp(i \Phi(f_0))$, since the time-reflection $\theta$ acts
trivially on these vectors. It remains to see that the restriction
to such vectors is \emph{onto}  $\cH$. Then we wish to prove
\bel{eq:cupt>0}
    \Quantization{(\cE_0)}
    \supset
    \Quantization{\lrpBig{\bigcup_{t>0} \cE_t}} \, .
\ee

First, let us see why \er{eq:cupt>0}, if true, finishes the proof.
We must show that $\bigcup_{t>0} \cE_t$ is dense in $\cE_+$. Of
course, $\cE_+$ is spanned by polynomials in classical fields
of the form
\[
    \Phi(f) =
    \int \Phi(x,t) f(x,t) \sqrt{\scrg} \, dx dt \, .
\]
Write the $t$ integral as a Riemann sum:
\begin{eqnarray}\label{ce_ti}
    \Phi(f)
    &=& \lim_{N \to \infty} \sum_{i=1}^N (\delta t)_i \ \Phi\lrp{(f_i)_{t_i}} ,
    \\
    \te{where} && \Phi\lrp{(f_i)_{t_i}} = \int \Phi(x,t_i) f_i(x) \sqrt{\scrg} \, dx ,
\end{eqnarray}
and where $f_i(x) = f(x, t_i)$.

Eqn.~\er{ce_ti} represents $\Phi(f)$ as a limit of linear combinations of
elements $\Phi(f_{t_i}) \in \cE_{t_i}$.
A similar argument applies to polynomials $A(\Phi)$ of classical fields,
and to $L^2$ limits of such polynomials. Thus
$\bigcup_{t>0} \cE_t$ is dense in $\cE_+$. Then \er{eq:cupt>0}
implies $\Quantization{(\cE_0)}$ is also dense in $\cE_+$.

Equation \er{eq:cupt>0} is proved by means of the following identity:
\bel{eq6.2.16}
    \< \hat A, : \exp( i\, \Phi(f_t)) : \HatOverSpace \>_{\cH}
    =
    \< \hat A, : \exp( i\, \Phi({f^t}_0)) : \HatOverSpace \>_{\cH}
\ee
where
\bel{def-of-ft}
    f^t\ :=\ (F^{-1/2} e^{-t \omega} F^{1/2}) f,
\ee
where $f$ is a function on $\Sigma$, and hence so is $f^t$. Thus
\[
    {f^t}_0(p,t') = \de(t') (F^{-1/2} e^{-t \omega} F^{1/2} f)(p)
    \quad \te{ for } \quad p \in \Sigma \, .
\]
To prove \er{eq6.2.16}, we first suppose $A = \no{ e^{i \Phi(g_s)} }$
where $g \in \cT_\Sigma$ and $s > 0$. Then
\begin{eqnarray}
    \< \hat A, \no{ \exp( i \Phi(f_t)) } \HatOverSpace \>_{\cH}
    &=&
    \< \no{ e^{i \Phi(\th g_s)} }, \no{ e^{i \Phi(f_t)} } \>_{\cE}
    \nonumber \\
    &=&
    \exp {\< \th g_s, C f_t \>_M}
    \nonumber \\
    &=&
    \exp \big\< g, F^{1/2} \f{e^{-(t+s) \omega}}{2\omega} F^{1/2} f
    \big\>_{\cK_\Sigma}
    \label{eq:IPH1}
\end{eqnarray}
where we have used localization (Theorem \ref{thm:dimreduc}) in the last line.

Computing the right side of \er{eq6.2.16} gives
\begin{eqnarray}
    \big\< \no{e^{i \Phi(\th g_s)}}, \no{ e^{i \Phi({f^t}_0)} } \big\>_{\cE}
    &=&
    \exp { \big\< \th g_s,\, C({f^t}_0) \big\>_M }
    \nonumber \\
    &=&
    \exp \big\< g, F^{1/2} \f{e^{-s \omega}}{2\omega} F^{1/2} f^t
\big\>_{\cK_\Sigma}
    \nonumber \\
    &=&
    \exp \big\< g, F^{1/2} \f{e^{-(t+s) \omega}}{2\omega} F^{1/2} f
\big\>_{\cK_\Sigma}
    = \er{eq:IPH1} .
    \nonumber
\end{eqnarray}
We conclude that
eqns.~\er{eq6.2.16}-\er{def-of-ft} hold true for $A = \no{ e^{i\Phi(g_s)}}$.
We then infer the validity of \er{eq6.2.16} for all $A$ in
the span of $\bigcup_{t>0} \cE_t$  by linear combinations and
limits.

Equation \er{eq6.2.16} says that for every
vector $v$ in a set that is dense in $\cH$, there exists $v' \in
(\cE_0)\HatOverSpace$ such that $L(v) = L(v')$ for any linear
functional $L$ on $\cH$. If $v \ne v'$ then we could find some
linear functional to separate them, so they are equal. Therefore
$(\cE_0)\HatOverSpace$ is a dense set, completing the proof of
Theorem \ref{thm:626}.
\end{proof}

Theorem \ref{thm:626} implies that the physical Hilbert space is
isometrically isomorphic to $\cE_0$, and to an $L^2$ space of the
Gaussian measure with covariance which can be found by the $t,s \to
0$ limit of \er{eq:IPH1}, to be:
\bel{modifiedcovariance}
    \cH = L^2\lrp{N_{d-1}^*,\ d\phi_{\fC}} \, ,
   \ \ \te{ where } \ \
   \fC = F^{1/2} \f{1}{2\omega}\, F^{1/2} \, ,
\ee
and $N_{d-1}$ denotes the nuclear space over the
$(d-1)$-dimensional slice. Compare \er{modifiedcovariance} with
\cite{GJ}, eqn.~(6.3.1). By assumption, 0 lies in the resolvent
set of $\omega$, implying that $\fC$ is a bounded, self-adjoint
operator on $\cK_\Sigma$.

\subsection{The $\vp$ bound}

Here we prove that an estimate known in constructive field theory as the
\emph{Glimm-Jaffe $\vp$ bound} (see \cite{GlimmJaffeIV}) is also true for curved
spacetimes. 

\begin{theorem}[$\vp$ bound] \label{Thm:PhiBound}
Let $T > 0$. There exists a constant $M$ such that
\bel{PhiBound}
    \exph{ \hat A, e^{-(H_0 + \varphi(h))T} \hat A }
    \leq
    \exp\Big( T \norm{h}_G^2 M \Big) \norm{\hat A}_{\H}^2,
\ee
where $\norm{h}_G = \<h, Gh\>^{1/2}$ and
$G$ is the resolvent of $Q$ at $-m^2$.
\end{theorem}

\begin{proof}
Apply the Schwartz
inequality (for the inner product on $\cH$) $n$ times, to obtain
\begin{eqnarray*}
    \exph{ \hat A, e^{-(H_0 + \varphi(h))T} \hat A}
    &\leq&
    \norm{\hat A}_{\H} \exph{ \hat A, e^{-2T(H_0 + \varphi(h))} \hat A}^{1/2}
    \\ &\leq&
    \norm{\hat A}_{\H}^{2-2^{-(n-1)}}
    \exph{ \hat A, e^{-2^n T (H_0 + \varphi(h))} \hat A }^{2^{-n}} .
\end{eqnarray*}
Apply the Feynman-Kac formula to the very last expression, to obtain
\[
	\exph{ \hat A, e^{-(H_0 + \varphi(h))T} \hat A}
	\leq
    \norm{\hat A}_{\H}^{2-2^{-(n-1)}}
    \expe{ \Th A, e^{- \int_0^{2^nT} \Phi(h,t) dt } U(2^n T)
    A}^{2^{-n}}.
\]
We cannot take the $n \to \infty$ limit at
this point, because the object depends on $A$. It suffices to establish the
desired result for $A$ in a dense subspace, so take $A \in L^4 \cap \cE_+$. We
now use the Schwartz inequality on $\cE$ as well as the fact that $\Th$ is
unitary on $\cE$, to obtain
\[
    \exph{ \hat A, e^{-(H_0 + \varphi(h))T} \hat A}
    \ \leq\
    \norm{\hat A}_{\H}^{2-2^{-(n-1)}}
    \norm{A}_\cE^{2^{-n}}
    \expe{ A, e^{2 \int_0^{2^nT} \Phi(h,t) dt } A}^{2^{-(n+1)}}
\]
Now H\"older's inequality with exponents $\f14 + \f14 + \f12 = 1$ implies
\begin{eqnarray} \nonumber
	\< \hat A, && e^{-(H_0 + \varphi(h))T} \hat A \>_{\cH}
	\\
    && \leq
    \norm{\hat A}_{\H}^{2-2^{-(n-1)}}
    \norm{A}_\cE^{2^{-n}}
    \norm{A}_{L^4}^{2^{-n}}
    \big(
			\int e^{4 \int_0^{2^nT} \Phi(h,t) dt} d\mu_0
	\big)^{2^{-n-2}} . \qquad
	\label{eq:integral}
\end{eqnarray}
Up to this point, the argument applies to a general measure
$d\mu$ on path space. Now assume that the measure is Gaussian. The function $f
= 4 h(\vec x) \chi_{[0, 2^n T]}(t)$ has the desirable property that $\Phi(f) = 4
\int_0^{2^nT} \Phi(h,t) dt$, so the Gaussian integral in \eqref{eq:integral}
equals $S(i f) = e^{\<f, Cf\>/2}$. Therefore,
\bel{eq:int2}
	\exph{ \hat A, e^{-(H_0 + \varphi(h))T} \hat A}
    \leq
    \norm{\hat A}_{\H}^{2-2^{-(n-1)}}
    \norm{A}_\cE^{2^{-n}}
    \norm{A}_{L^4}^{2^{-n}}
    S(if)^{2^{-n-2}} .
\ee

For $H_1$ and $H_2$ self-adjoint operators with $0 \leq H_1 \leq H_2$, we have
$(H_2+a)^{-1} \leq (H_1+a)^{-1}$ for any $a>0$.
By theorem \ref{thm:Qsymmetric}, $-Q \geq 0$, so take $H_1 = -Q$, and $H_2 =
-(1/F) \d_t^2 - Q$. We conclude\footnote{Compare this with the analogous
estimate valid in $\R^d$, $C \leq ( - \del_{\vec x}^2 + m^2 )^{-1}$, which may
be proved by a Fourier transform of the resolvent
kernel.}
\[
    C \,=\, (-\De + m^2)^{-1} \leq (-Q + m^2)^{-1} \, \equiv \,  G .
\]
Since $\ker(G) = \{ 0 \}$, $G$ determines a norm $\norm{h}_G = \<h,
Gh\>^{1/2}$. Then
\[
    S(i f) \leq e^{8 \<h, Gh\> 2^n T} =
    e^{2^{n+3} T \norm{h}_{G}^2} \, .
\]
Raising this to the power $2^{-n-2}$, and taking the
$n \to \infty$ limit we see that the
factors $\norm{A}_\cE^{2^{-n}} \norm{A}_{L^4}^{2^{-n}}$
approach 1, and thus \eqref{eq:int2} becomes:
\[
    \exph{ \hat A, e^{-(H_0 + \varphi(h))T} \hat A }
    \leq
    e^{2 T \norm{h}_{G}^2} \norm{\hat A}_{\H}^2 .
\]
This establishes \er{PhiBound}, completing the proof of Theorem
\ref{Thm:PhiBound}.
\end{proof}

%%%%%%%%%%%%%%%%%%%%%%%%%%%%%%%%%%%%%%%%%%%%%%%%%%%%%%%%%%%%
\subsection{Fock representation for time-zero fields}
%%%%%%%%%%%%%%%%%%%%%%%%%%%%%%%%%%%%%%%%%%%%%%%%%%%%%%%%%%%%

To obtain a Fock representation of the \emph{time-zero fields} we
mimic the construction of \cite[\S\ 6.3]{GJ} with the
covariance \er{modifiedcovariance}.

To simplify the constructions in this section, we assume the form
$ds^2 = dt^2 + G_{\m\n} dx^\m dx^\n$ and $F = 1$. Then $Q =
\De_\Sigma$, the Laplacian on the time-zero slice, and $\mu =
(-\De_\Sigma + m^2)^{1/2}$. The set of functions
$h \in L^2(\Sigma)$ such that ${\mu}^p h \in L^2(\Sigma)$ is
precisely the Sobolev space $H_p(\Sigma)$, which is also the set of
$h$ such that $\fC^{-p} h \in L^2$. Sobolev spaces satisfy the
reverse inclusion relation $p \geq q \implies H_q \subseteq H_p$.
Also $\fC^q f \in H_p \iff f \in H_{q-p}$.

This allows us to determine the natural space of test functions
for the definition of the Fock representations:
\begin{eqnarray*}
    a(f) &=& \f12 \phi\lrp{\fC^{-1/2}f} + i \pi\lrp{\fC^{1/2} f}
    \\
    a^*(f) &=& \f12 \phi\lrp{ \fC^{-1/2} f} - i\pi\lrp{\fC^{1/2}f} \, .
\end{eqnarray*}
In particular, if the natural domain of $\phi$ is $H_{-1}$ as
discussed following eqn.~\er{eqn:first-order}, then $f$ must lie
in the space where $\fC^{-1/2}f \in H_{-1}$, i.e. $f \in H_{1/2}$.

\section{Conclusions and Outlook}

We have successfully generalized Osterwalder-Schrader quantization and
several basic results of constructive field theory to the setting of static
spacetimes.

Dimock \cite{Dimock84} constructed an interacting
$\mathcal{P}(\varphi)_2$ model with variable coefficients, with
interaction density $\rho(t,x) :\varphi(x)^4:$, and points out that
a Riemannian $(\varphi^4)_2$ theory may be reduced to a Euclidean
$(\varphi^4)_2$ theory with variable coefficients. However, the main
constructions of \cite{Dimock84} apply to the Lorentzian case and for curved
spacetimes no analytic continuation between them is known. Establishing the
analytic continuation is clearly a priority. Also, there
are certain advantages to a perspective which remembers the spacetime structure;
for example, in this picture the procedure for quantizing spacetime
symmetries is more apparent.

In the present paper we have not treated the case of a
non-linear field, though all of the groundwork is in place. Such
construction would necessarily involve a generalization of the
Feynman-Kac integral \er{fk} to curved space, and would have
far-reaching implications, and one would like to establish
properties of the particle spectrum for such a theory.

The treatment of symmetry in this paper is only preliminary. We have isolated
two classes of isometries, the reflected and reflection-invariant
isometries, which have well-defined quantizations. We believe that this
construction can be extended to yield a unitary representation of the isometry
group, and work on this is in progress. This, together with suitable extensions
of section \ref{sec:SubgroupsOfIsometries} could have implications for
the representation theory of Lie groups, as is already the case for the
geometric quantization of classical Hamiltonian systems.

The treatment of variation of the metric in section \ref{sec:VariationOfMetric}
is also preliminary; it does not cover the full class of static spacetimes.
Geroch \cite{Geroch:1969} gave a rigorous definition of the limit of a family of
spacetimes, which formalizes the sense in which the Reissner-Nordstr\"{o}m black hole becomes 
the Schwarzschild black hole in the limit of vanishing charge. It would be interesting to
combine the present framework with Geroch's work to study rigorously the
properties of the quantum theory under a limit of spacetimes.

Another direction is to isolate specific spacetimes
suggested by physics which have high symmetry or other special
properties, and then to extend the methods of constructive field
theory to obtain mathematically rigorous proofs of such
properties. Several studies along these lines
exist \cite{BEM02,Ishi-Wald}, but there is much more to be done. We hope that
the Euclidean functional integral methods developed here may facilitate further
progress. Rigorous analysis of thermal properties such as Hawking radiation
should be possible. Given that new mathematical methods are available which
pertain to Euclidean quantum field theory in AdS, a complete, rigorous
understanding of the holographically dual theory on the boundary of $AdS$
suggested by Maldacena \cite{Malda,Witten,GKP,MaldaReview} may be within reach
of present methods.

Constructive field theory on flat spacetimes has
been developed over four decades and comprises thousands of
published journal articles. Every statement in each of those
articles is either: (i) an artifact of
the zero curvature and high symmetry of $\R^d$ or $\mathbb{T}^d$
or (ii) generalizable to curved spaces with less
symmetry. The present paper shows that the Osterwalder-Schrader
construction and many of its consequences are in class (ii).
For each construction in class
(ii), investigation is likely to yield non-trivial connections
between geometry, analysis, and physics.

    \vskip 0.2 in

\subsubsection*{Acknowledgements}
We would like to thank Jonathan Weitsman and Joachim Krieger for
interesting discussions, and Jon Dimock for his earlier work
\citep{Dimock84,Dimock:2003an} which sparked our interest in these
models.

\appendix

\section{Euclidean Anti-de Sitter and its Analytic Continuation}
\label{sec:AdSGreensFunctions}

The Green's function $G$ on a general curved manifold is the
inverse of the corresponding positive transformation, so it
satisfies
\bel{AdsGreenEq}
    (\Delta - \mu^2) G = -g^{-1/2} \delta\;,
\ee
where $G(p,q)$ is a function of two spacetime points. By
convention $\Delta$ acts on $G$ in the first variable, and
$\delta$ denotes the Dirac distribution of the geodesic distance
$d = d(p,q)$. Translation invariance implies that $G$ only depends
on $p$ and $q$ through $d(p,q)$. We note that solutions of the
homogeneous equation $(\Delta - \mu^2) \phi = 0$ may be recovered
from the Green's function. Conversely, we may deduce the Green's
function by solving the homogeneous equation for $d > 0$ and
enforcing the singularity at $d = 0$.

The equation \er{AdsGreenEq} for the Green's function takes a
simple form in geodesic polar coordinates on $\hyp{n}$ with $r = d = $ geodesic distance; the Green's function has no dependence on the angular variables and the radial equation yields
\bel{green1}
    \lrp{ \d_r^2 + (n-1) \coth(r) \d_r - \mu^2 } G(r) = -\delta(r) \,
    .
\ee
We find it convenient to write the homogeneous equation in terms
of the coordinate $u = \cosh(r)$. When $u \ne 1$, \er{green1}
becomes
\bel{green2}
    (\Delta - \mu^2) G(u)
    =
    -(1-u^2) G''(u) + n u G'(u) - \mu^2 G(u) = 0 \, .
\ee
For $n = 2$ and $\mu^2 = \nu(\nu+1)$, eqn.~\er{green2} is
equivalent to Legendre's differential equation:
\bel{Legendre}
    (1-u^2) Q_\nu''(u) -2u Q_\nu'(u) + \nu(\nu+1) Q_\nu(u) = 0 \, .
\ee
Note that \er{Legendre} has two independent solutions for each
$\nu$, called Legendre's $P$ and $Q$ functions, but the $Q$
function is selected because it has the correct singularity at $r
= 0$. Thus
\bel{G2}
    G_2(r; \mu^2)
    =
    \f{1}{2\pi} Q_\nu(\cosh r),
    \ \ \te{ where } \ \
    \nu = -\f12 + \lrp{ \mu^2 + \f14}^{1/2} \, .
\ee
The case $\mu^2 = 0$ is particularly simple; there the Legendre
function becomes elementary:
\bel{G20}
    G_2(r; 0) = -\f1{2\pi} \ln\lrp{ \tanh \f{r}{2} }
    =
     \f1{2\pi} Q_0(\cosh r) \, .
\ee
For $n = 3$, one has
\bel{G3}
    G_3(r; \mu^2)
    =
    \f1{4\pi}
    \f{e^{\pm r \sqrt{\mu^2+1}}}{\sinh(r)} \, .
\ee Finally, we note that the analytic continuation of \er{G2} gives
the Wightman function on $AdS_2$. The real-time theory on Anti-de
Sitter, including its Wightman functions, were discussed by Bros
et al. \cite{BEM02}. In particular, our equation \er{G2}
analytically continues to their equation (6.8).

Given a complete set of modes, one may also calculate the Feynman
propagator by using the relation  $i G_F(x,x') = \< \, 0 \, | \,
T\{ \phi(x) \phi(x') \} \, | \, 0 \, \>$ and performing the mode
sum explicitly as in \cite{BL}; the answer may be seen to be
related to the above by analytic continuation. Here, $T$ denotes
an $AdS$-invariant time-ordering operator. A good general
reference is the classic paper \cite{ais78}.

\bibliographystyle{plain}

\def\cprime{$'$}

\end{document}